# Monte Carlo Simulation of Lane-Emden Type Equations Arising in Astrophysics

Samah H. El-Essawy[1], Mohamed I. Nouh[1], Ahmed A. Soliman[2], Helal I. Abdel Rahman[1], Gamal A. Abd-Elmougod[3]

[1]Astronomy Department, National Research Institute of Astronomy and Geophysics, 11421 Helwan, Cairo, Egypt
[2]Department of Mathematics, Faculty of Science, Sohag University, Sohag, 82524, Egypt
[3]Department of Mathematics, Faculty of Science, Damanhour University, Damanhour 22511, Egypt

**mohamed.nouh@nriag.sci.eg**

**Abstract:** Monte Carlo (MC) method played an essential role in many areas of human activity and has found application in many branches of physical science. This paper proposes a computational technique based on MC algorithms to solve Lane-Emden (LE) type equations. We analyze four LE equations arising in astrophysics: the positive and negative indices of the polytropic gas spheres, the isothermal gas sphere, and the white dwarf equation. We calculated eleven models (i.e., eleven LE equations) of the positive index polytropes, nine for the negative index polytrope, the isothermal gas sphere, and the white dwarf equation. Comparing the MC and numerical/analytical models gives good agreement for the four LE equations under study.

## 1. Introduction

Lane-Emden (LE) equations play an essential role in the theory of star structure and evolutions, thermodynamics, modeling of galaxy clusters, and many other physics, chemistry, and engineering topics. LE equation can be referred to as a Poisson equation that relates the gradient of the gravitational potential of the self-gravitating, symmetric gas sphere with its density and radius. There is an exact solution of LE equations for only limited values of the polytropic indices, so numerical or analytical techniques might solve these equations.

A preliminary study on the LE equations (polytropic and isothermal) was undertaken by astrophysicists Lane (1870) and Emden (1907). Since this time, a variety of methods have been used to estimate analytical as well as numerical solutions of the LE equations, such as Runge-Kutta type methods, Horedt (1986), Adomian decomposition methods, Shawagfeh (1993), and Wazwaz (2001), Homotopy perturbation methods, Chowdhury and Hashim (2007) and Yildirim and Özis (2007), Series expansion, Ramos (2008), an accelerated power series method Nouh (2004), Variational iteration method, Dehghan and Shakeri (2008), Differential transform method, Mukherjee (2011) and many other methods.

Various novel computing approaches for increasingly complicated problems have emerged fast in recent decades, such as the Genetic Algorithm (e.g., Ge et al., (2008)), Lattice Boltzmann method (e.g., Zhang et al., (2003)), Ant colony Algorithm (e.g., Cao and Guo, (2011)), Artificial neural networks (e.g., Morawski and Bejger (2020), Abdel-Salam et al. (2021), Azzam et al. (2021), Nouh et al. (2021).

The Monte Carlo method is mainly a series of statistical techniques used to either get solutions of function when it can't analytically be solved or evaluate the estimated value of a parameter or a function of a specific distribution. In general, any problem can be solved under two techniques of the Monte Carlo method (simulation and integration). In contrast to traditional methods, primarily used to solve problems in specific fields, the Monte Carlo method has been studied as an independent method over the last century as science, technology, and computers have advanced. In the late 1960s, Monte Carlo calculations entered the astrophysics stage, for example, with the works by Auer (1968), Avery and House (1968), and Magnan (1968, 1970). One of the most powerful Monte Carlo algorithms is the Markov chains algorithm (MC). MC method (see, for example, Hestroffer (2012), Mede & Brandt (2014)) is currently widely utilized in exoplanet research, for astrometric orbits (Tuomi & Kotiranta (2009), Otor et al. (2016)), for visual binary orbits Mendez et al. (2017) and other topics.

Compared to deterministic or "single point estimate" analysis, Monte Carlo simulation has various benefits. Results demonstrate both what may occur and how probable each possibility is. Analysis may determine precisely which inputs had values combined when specific outcomes occurred. Based on the empirical experiments, more samples generate as the obtained solution gets closer to the exact solution. A few published papers investigate the Monte Carlo method to provide solutions to linear and non-linear differential equations. In 2011, Zhong and Tian (2011) reported a new way of the Monte Carlo method to solve the initial value problem of ordinary differential equations. Akhtar et al. (2015) suggested a new algorithm of the Monte Carlo method that gives accurate solutions to different types of ordinary differential equations. This algorithm efficiently refines the complications in the scheme of Zhong and Tian (2011). Recently, Uslu and Sari (2020) discussed a Monte Carlo-based stochastic approach for different systems of the Lotka- Volterra equation. No studies have investigated Monte Carlo integration for accurate numerical solutions for the LE equations.

Although traditional numerical methods like the Rung-Kutta method could be used to solve LE equations, numerous numerical techniques have been proposed to do so, including neural networks (Ahmad et al. 2017, Nouh et al. 2021), genetic algorithms (Ahmad et al., 2016), and the pattern search optimization technique (Lewis et al. 2000). In this study, we introduce four LE equations' numerical solutions utilising the MC technique. We shall study four LE equations, including the polytropic LE equation, the isothermal LE equation, and the white dwarf equation. The obtained results from the MC approach will be compared to the exact and numerical solutions to demonstrate their efficiency and accuracy. The paper's overall structure builds on four sections, including this introductory section. The second section briefly overviews the derivation of the LE type equations and white dwarfs' equations. The third section has implemented the producers of Monte Carlo integration used for this study. The fourth section analyses the numerical results. The fifth section gives the conclusion of the paper.

## 2. Lane-Emden Equations

This section briefly derives the LE equations (polytropic and isothermal gas spheres) and the white dwarf equation.

### 2.1. Polytropic gas sphere

A fundamental equation in the theory of stellar structure is the LE equation of positive polytropic index $n$. According to the rules of thermodynamics and the mutual attraction of its molecules, the equation represents the temperature variation of a spherical gas cloud. The Polytropic model is a simplified method of equations that relates the pressure and density of the star by the polytropic equation of state

$$P(r) = K\,\rho^{1+\frac{1}{n}}(r), \tag{1}$$

with some constant $K$, radius $r$, and the polytropic index $n$.

The polytropic LE equation was the equation of mass continuity and hydrostatic equilibrium as follows

$$\frac{dm(r)}{dr} = 4\pi r^2 \rho(r), \tag{2}$$

$$\frac{dP(r)}{dr} = -G\frac{m(r)}{r^2}\rho(r). \tag{3}$$

Combining equation (2) with equation (3), then multiply by $\frac{r^2}{\rho(r)}$ and differentiate for $r$, holds

$$\frac{d}{dr}\left(\frac{r^2}{\rho(r)}\frac{dP(r)}{dr}\right) = -G\frac{dm(r)}{dr}. \tag{4}$$

Substitute equation (2) into equation (4), and we get

$$\frac{1}{r^2}\frac{d}{dr}\left(\frac{r^2}{\rho(r)}\frac{dP(r)}{dr}\right) = -4\pi G \rho(r). \tag{5}$$

For simplicity, we transform equation (5) into a dimensionless form by defining dimensionless variables $\theta$ and $r$

$$\begin{aligned}\rho(r) &= \rho_c(r)\,\theta^n(r), \\ r &= \alpha\xi.\end{aligned} \tag{6}$$

Inserting equation (6) into equation (1) leads to

$$\begin{aligned}P(r) &= K\,\rho_c^{1+\frac{1}{n}}(r)\theta^{n+1}(r) = P_c(r)\theta^{n+1}(r) \\ P_c(r) &= K\,\rho_c(r).\end{aligned} \tag{7}$$

Then Eq. (5) becomes

$$\frac{1}{\alpha^2}\left[\frac{(n+1)K}{4\pi G \rho_c^{1-\frac{1}{n}}(r)}\right]\frac{1}{\xi^2}\frac{d}{d\xi}\left(\xi^2\frac{d\theta(\xi)}{d\xi}\right) = -\theta^n(\xi), \tag{8}$$

with some constant $\alpha^2$ ($\alpha$ in centimeters) in squared brackets, so

$$\alpha = \left[\frac{(n+1)K}{4\pi G \rho_c^{1-\frac{1}{n}}(r)}\right]^{\frac{1}{2}}. \tag{9}$$

Combining equation (8) with equation (9) yields

$$\frac{1}{\xi^2}\frac{d}{d\xi}\left(\xi^2\frac{d\theta(\xi)}{d\xi}\right) = \mp\theta^n(\xi). \tag{10}$$

The right-hand side of equation (10) takes the minus sign when the polytropic indices $-1 < n < \infty$, while the plus sign holds when $-\infty < n < -1$.

Now, the solution of equation (10) holds under the initial conditions

$\frac{d\theta(\xi)}{d\xi} = 0, \theta(\xi) = 1$ at $\xi = 0$.

To solve equation (10), various analytical and numerical techniques have been applied recently. The equations' singularity causes the primary challenge at $\xi = 0$. There is an exact solution for some values of the polytropic index $n$, as follows

$$n = 0 \to \theta(\xi) = 1 - \frac{\xi^2}{6}$$

$$n = 1 \to \theta(\xi) = \frac{\sin\xi}{\xi} \tag{11}$$

$$n = 5 \to \theta(\xi) = \frac{1}{\left(1 + \frac{\xi^2}{3}\right)^{\frac{1}{2}}}$$

For other values of $n$, $\theta(\xi)$ can be computed by numerical or analytical methods.

## 2.2 The Isothermal gas sphere

For an isothermal gas sphere, set $n \to \infty$ and regard a temperature $T$ as a constant value. Combining equation (5) with $P(r) = K\rho(r)$, where $K = \frac{kT}{\mu H}$, gets

$$\frac{K}{r^2}\frac{d}{dr}\left(\frac{r^2}{\rho(r)}\frac{d\rho(r)}{dr}\right) = -4\pi G\rho(r). \tag{12}$$

It is convenient to rewrite equation (12) as,

$$\frac{K}{r^2}\frac{d}{dr}\left(r^2\frac{d}{dr}\log\rho(r)\right) = -4\pi G\rho(r). \tag{13}$$

Set $\rho(r) = \rho_c(r)e^{-\varphi}$ and $r = \alpha\xi$. Under the previous assumptions, equation (13) becomes,

$$\frac{K}{4\pi G\rho_c(\xi)}\frac{1}{\alpha^2\xi^2}\frac{d}{d\xi}\left(\xi^2\frac{d\varphi(\xi)}{d\xi}\right)=e^{-\varphi(\xi)}, \tag{14}$$

where $\alpha$ can be chosen to have the formulae $\alpha=\left[\frac{K}{4\pi G\rho_c(\xi)}\right]^{\frac{1}{2}}$.

We write equation (14) for short, as

$$\frac{1}{\xi^2}\frac{d}{d\xi}\left(\xi^2\frac{d\varphi(\xi)}{d\xi}\right)=e^{-\varphi(\xi)}. \tag{15}$$

Only numerical solutions are held when $\varphi(\xi)$ is subject to the conditions

$$\frac{d\varphi(\xi)}{d\xi}=0, \varphi(\xi)=0 \; at \; \xi=0.$$

## 2.3 White Dwarf equation

Chandrasekhar (1958) derived an initial value problem ordinary differential equation, namely Chandrasekhar white dwarf equation.

Taking pressure $P$ and density $\rho$ as

$$\begin{aligned}P(x)&=Af(x),\\ \rho(x)&=Bx^3,\end{aligned} \tag{16}$$

where

$$\begin{aligned}A&=\frac{\pi m_e^4 c^5}{3h^3}=6.02\times 10^{21},\\ B&=\frac{8\pi\mu_e m_p m_e^3 c^3}{3h^3}=9.82\times 10^8,\\ f(x)&=x(2x^2-3)(x^2+1)^{\frac{1}{2}}+3\sinh^{-1}x.\end{aligned} \tag{17}$$

Combining equation (16) with equation (5) gets

$$\frac{A}{B}\frac{1}{r^2}\frac{d}{dr}\left(\frac{r^2}{x^3}\frac{df(x)}{dr}\right)=-4\pi GBx^3. \tag{18}$$

$$\frac{1}{x^3}\frac{df(x)}{dr} = \frac{8x}{(x^2+1)^{\frac{1}{2}}}\frac{dx}{dr} = 8\frac{d\sqrt{x^2+1}}{dr}. \tag{19}$$

Now, equation (18) becomes

$$\frac{1}{r^2}\frac{d}{dr}\left(r^2 \frac{d\sqrt{x^2+1}}{dr}\right) = -\frac{\pi G B^2}{2A}x^3. \tag{20}$$

Set $y^2 = x^2 + 1, r = \mu\eta, and\ y = y_0\psi$, yields

$$\frac{1}{y_0^2}\left(\frac{2A}{\pi G B^2}\right)\frac{1}{\mu^2\eta^2}\frac{d}{d\eta}\left(\eta^2 \frac{d\psi}{d\eta}\right) = -\left(\psi^2 - \frac{1}{y_0^2}\right)^{\frac{3}{2}}. \tag{21}$$

Then equation (21) can be rewritten in the formulae

$$\frac{1}{\eta^2}\frac{d}{d\eta}\left(\eta^2 \frac{d\psi}{d\eta}\right) = -(\psi^2 - C)^{\frac{3}{2}}. \tag{22}$$

With some fixed values of $\mu = \frac{1}{y_0}\left(\frac{2A}{\pi G B^2}\right)^{\frac{1}{2}}$ and $C = \frac{1}{y_0^2}$

The existence of solutions $\psi(\eta)$ are held under initial conditions as

$$\frac{d\psi(\eta)}{d\eta} = 0, \psi(\eta) = 1\ at\ \eta = 0.$$

When $\eta$ tends to $\infty$, then $\psi(\eta_\infty) = \sqrt{C}$ and the range of $\psi(\eta)$ becomes

$$\sqrt{C} \le \psi(\eta) \ge 1.$$

The constant $C$ takes values $C = 0.01 - 0.8$.

Equation (22) is of LE type where $f(y) = (y^2 - C)^{\frac{3}{2}}$. If $C = 0$, equation (22) reduces to LE equation, Equation (10) of positive polytropic index $n = 3$.

## 3. The MC Method

### 3.1. The MC integration

The integral of the function $f(x)$ over an interval width $(b - a)$ is given by

$$I = \int_a^b f(x)dx = (b-a)<f>,$$

where $<f>$ is the average value of the function $f$. We could then calculate the integral if we had some independent technique for estimating the integrand's average value. Assume we have a list of random numbers, $x_i$, where $i = 1, 2, \ldots, N$ that are evenly distributed between $a$ and $b$, then the average value of the function is given by

$$<f>_N = \frac{1}{N}\sum_{i=1}^{N} f(x_i).$$

In the following subsections, we shall solve an explicit and implicit form of ODEs using the MC method.

### 3.2. Explicit ODE

Let us assume explicit ODE is written in general form as

$$\frac{dy(x)}{dx} = f(x), \quad with \;\; y(x_0) = y_0. \tag{23}$$

Firstly, integrate both sides of equation (23) concerning $x$ over some specified interval and gets

$$y(x) = y(x_0) + \int_{x_0}^{x} dt\, f(t). \tag{24}$$

Secondly, split the limit of integration into small, discrete chunks, as

$$y(x) = y(x_0) + \sum_{i=1}^{N} \int_{x_{i-1}}^{x_i} dt\, f(t). \quad where\; x = x_N \tag{25}$$

Thirdly, the definite integral on the right-hand side can be estimated using the MC method by drawing $M$ samples of $x$ due to uniform distribution U $(x_{i-1}, x_i)$

$$y(x) = y(x_0) + \sum_{i=1}^{N} \left[\frac{x_i - x_{i-1}}{M} \sum_{j=1}^{M} f(x_j)\right]. \tag{26}$$

Finally, we apply the Markov chains concept (which states that any future generated value depends only on the previous ones in the same sample) to evaluate the value of $y(x)$ for every discrete chunk.

$$y(x_i) = y(x_{i-1}) + \frac{x_i - x_{i-1}}{M} \sum_{j=1}^{M} f(x_j). \tag{27}$$

### 3.3. Implicit ODE

Set implicit ODE simply in the general form as,
$$\frac{dy(x)}{dx} = f(y(x), x), \quad with \quad y(x_0) = y_0. \tag{28}$$
In an explicit ODE, solution $y(x)$ of equation (28) can be estimated as,
$$y(x_i) = y(x_{i-1}) + \frac{x_i - x_{i-1}}{M} \sum_{j=1}^{M} f(y(x_{i-1}), x_j). \tag{29}$$

### 3.4. The MC algorithm for solving LE equations

For simplicity, we generally write the LE equations as

$$\frac{d^2\theta(\xi)}{d\xi^2} = -\frac{2}{\xi^2}\frac{d\theta(\xi)}{d\xi} + f(\theta(\xi), \xi). \tag{30}$$

The solution $\theta(\xi)$ can be integrated with the following steps:

1. Transform equation (30) into a first-order ordinary differential equation, define $\theta(\xi) = y_1(\xi)$, and $\frac{dy_1(\xi)}{d\xi} = y_2(\xi)$. Then equation (30) can be reduced to two first order differential equations as,

$$\frac{dy_1(\xi)}{d\xi} = y_2(\xi).$$
$$\frac{dy_2(\xi)}{d\xi} = -\frac{2}{\xi} y_2(\xi) + f(y_1(\xi), \xi) = g(y_2(\xi), y_1(\xi), \xi). \tag{31}$$

2. Split the interval $\left[0, \xi_f\right]$, where $\xi_f$ is the first zero of $y_1(\xi)$, into small, discrete chunks with $\Delta\xi = 10^{-3}$.
3. Generate a random sample from a uniform distribution from $\xi_i$ to $\xi_{i+1}$, with sample size $M = 1e^6$.
4. Initialize $\xi \to 0, y_1(\xi) \to 1 \; and \; y_2(\xi) \to 0$.
5. Evaluate $y_2(\xi_{i+1}) = y_2(\xi_i) + \frac{\xi_{i+1} - \xi_i}{M}\sum_{k=1}^{M} g(y_2(\xi_i), y_1(\xi_i), \xi_k)$.
6. Compute solution $y_1(\xi_{i+1}) = y_1(\xi_i) + \Delta\xi \times y_2(\xi_{i+1})$.
7. Set in the second step $\xi_{i+1} = \xi_i + \Delta\xi$.
8. Repeat steps 3-7, until $\xi_{i+1} = \xi_f$.

## 4. Results

To solve the four LE differential equations (equations (10), (15), and (22)), we elaborated $R$ codes that used the MC technique presented in section 3. For each case, a total of 1000000 random samples are used. The four differential equations are solved numerically using the $R$ package function rkMethod, based on the Runge-Kutta technique (RK) for solving ordinary differential equations. An odd, spaced interval with increments of 0.001 was chosen. After applying the method to the problem, the predicted MC results were compared to the RK results. We listed in Appendix A the numerical values obtained for the Emden functions for each of the four studied equations. In the following subsections, we analyze the behaviors of the results in more detail for each equation.

### 4.1 Positive index polytropes

For the positive polytrope (LE equation with $-1 < n < \infty$), the exact solutions are used to compute the Emden functions for $n = 0, 1$, and 5. The numerical or analytical solutions are used for the remaining polytropic indices due to the lack of an exact solution. The numerical results are listed in Tables A1-A12. In Figure 1, we plotted the results for the polytropic indices $n = 0, 1,$ and $5$, the MC, and the exact solutions are displayed in various colors to measure calculation; it is seen that they overlap and cannot be distinguished.

Besides the numerical integration, many methods are proposed to solve LE equations. We shall compare the MC polytropic models with the accelerated power series (APS) models developed by Nouh and Abdel-Salam (2018). Figure 2 plots the Emden function calculated for the polytropic indices $n = 0.5, 1.5, 2, 2.5,$ and 3 by the MC, RK, and APS, respectively. The results of the three methods are plotted with different colors; however, one can barely distinguish between them.

Another way to confirm the findings is to compare and validate the zeroth of the Emden function obtained using the MC algorithm ($\xi_1(MC)$) with that of the exact and RK solutions ($\xi_1$). The results for various polytropic indices are shown in Table 1. The zeros of the Emden function calculated using the proposed MC technique are shown in the second column, while the zeros computed from the exact or RK solutions are represented in the third column. The relative errors in the fourth column revealed good agreement between the two solutions, with a maximum value of 0.04%.

Table 1: The zeroth of the Emden function of the positive polytrope computed by the MC, exact/ RK solutions.

| $n$ | $\xi_1$(MC) | $\xi_1$ | Relative error % |
|---|---|---|---|
| 0 | 2.449 | 2.450 | 0.040 |
| 0.5 | 2.750 | 2.750 | 0 |
| 1 | 3.141 | 3.142 | 0.031 |
| 1.5 | 3.654 | 3.654 | 0 |
| 2 | 4.352 | 4.353 | 0.022 |
| 2.5 | 5.354 | 5.355 | 0.018 |
| 3 | 6.896 | 6.897 | 0.014 |
| 3.5 | 9.538 | 9.536 | 0.020 |
| 4 | 14.973 | 14.971 | 0.013 |
| 4.5 | 31.840 | 31.836 | 0.012 |

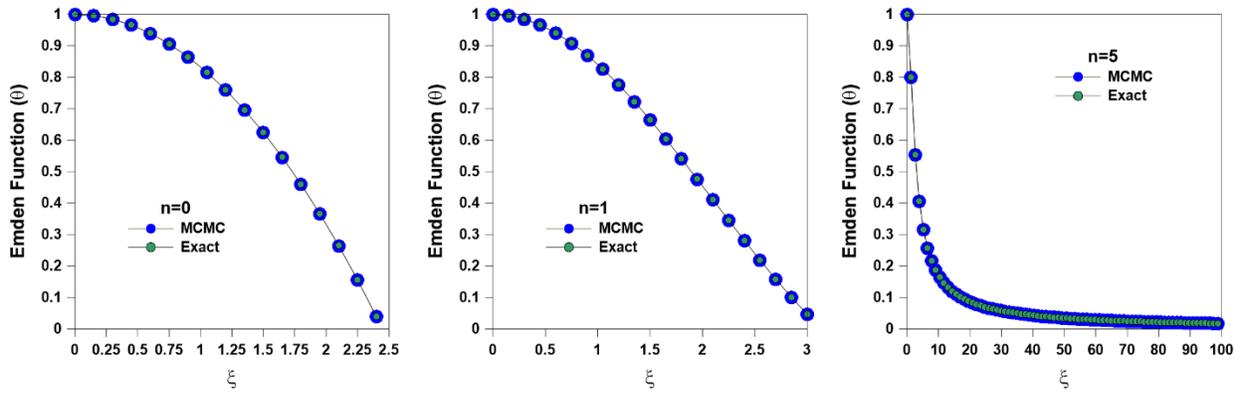

Figure 1: Comparison between the MC and the exact solutions computed for the polytropic indices $n = 0, 1,$ and $5$.

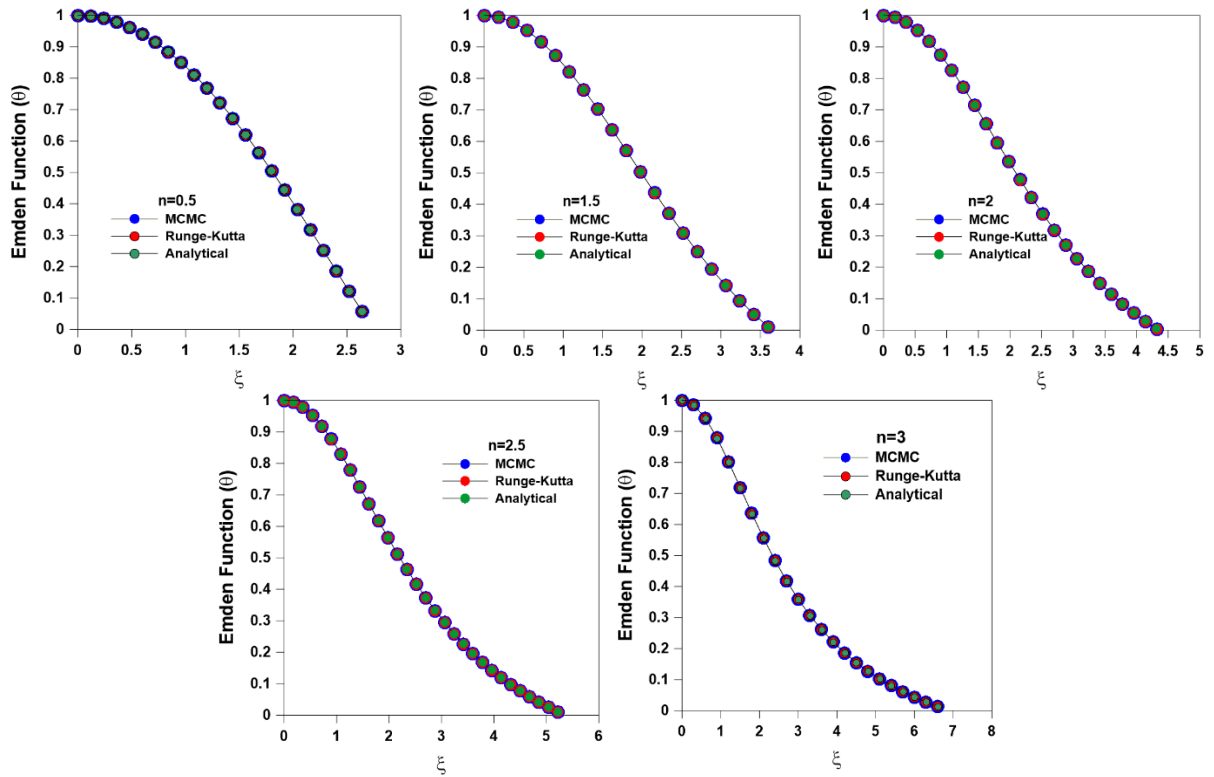

Figure 2: Comparison between the MC, numerical, and analytical solutions computed for the polytropic indices $n = 0.5, 1.5, 2, 2.5,$ and $3$.

## 4.2. Negative index polytropes

The term "negative polytropic index" refers to a process where heat and work are transferred through a system's boundaries concurrently. Such spontaneous processes break the Second Law of thermodynamics. These particular instances are exploited in thermal interaction for some astrophysical applications and chemical energy. There are no exact solutions for the negative polytrope (i.e., LE with $-\infty < n < -1$), so numerical or analytical methods could be used to solve it. Tables A13-A19 listed the numerical results for the polytropic indices $n = -1.01, -1.5, -2, -3, -4, -5,$ and $-10$.

Figure 3 plots the distributions of the Emden function with the dimensionless parameter $\xi$. As obtained for the polytropes with positive indices, there is good agreement between the MC and the RK solutions. Values of the Emden functions for the polytropic indices $n = -0.95\ \&\ -0.99$ are unity in the center and decrease toward the surface of the gas sphere. The situation for $n < -1$ is different than that of the polytropes with positive indices; the Emden functions increase monotonically, and there is no zeroth that the Emden function could determine.

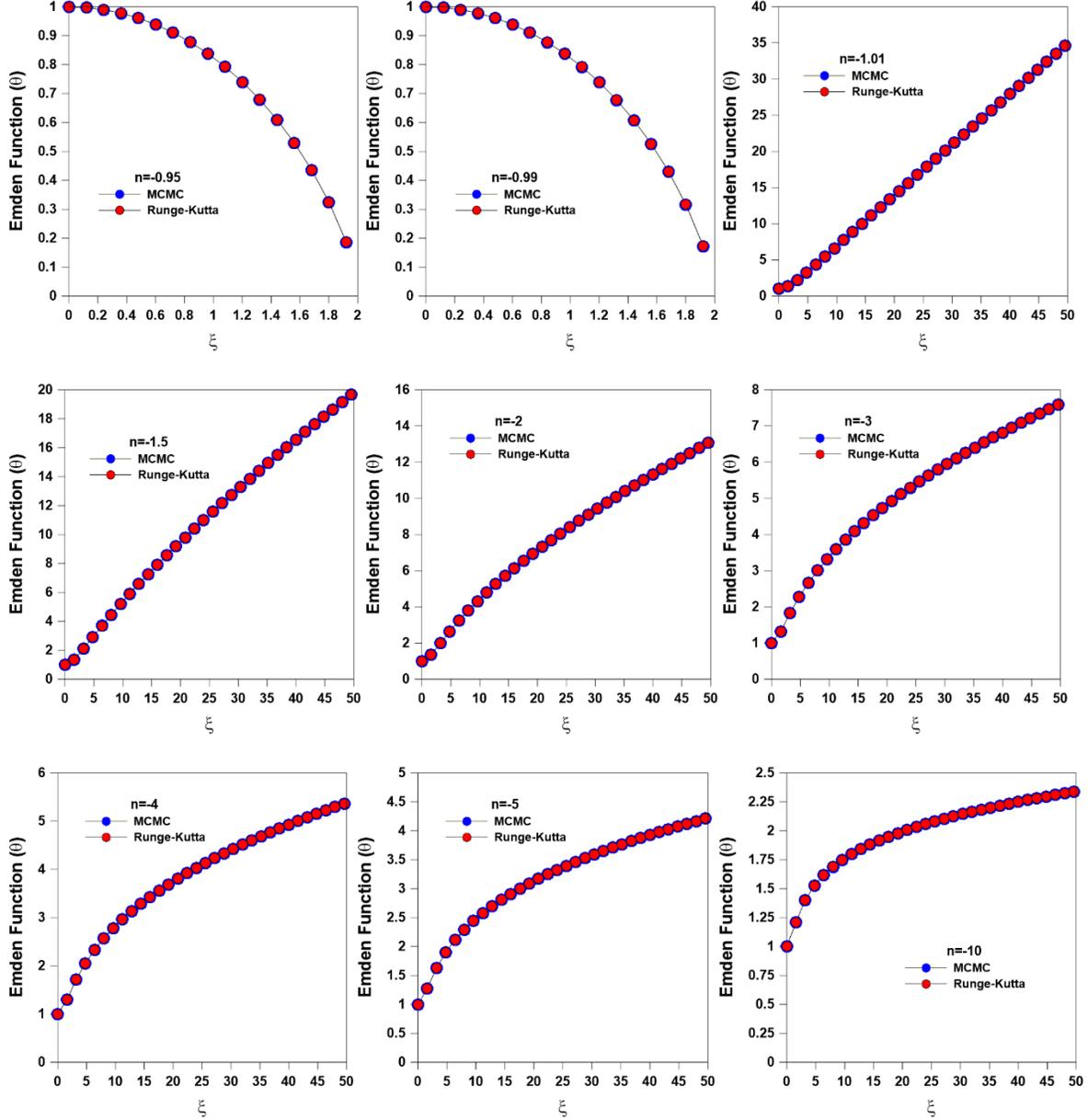

Figure 3: Comparison between the MC and numerical solutions computed for the polytropic indices $n = -0.95, -0.99, -1.01, -1.5, -2, -3, -4, -5, -10$.

## 4.3 The Isothermal gas sphere

The isothermal gas sphere (also called the second type LE equation) is a specific LE equation with $n \to \infty$. It is often used to represent various astrophysical issues, including the star, star cluster, and galaxy formation. This equation has no exact solution and is solved by numerical or analytical methods. The numerical result is presented in Table A20. In Figure 4, we compare the MC solution with the RK solution. The calculations of the MC and RK Emden functions are limited to the upper limit of $\xi = 35$, where this range is that of an isothermal sphere on the brink of gravo-thermal collapse, Hunter (2001). The comparison indicates good agreement with a maximum relative error of $0.00005$.

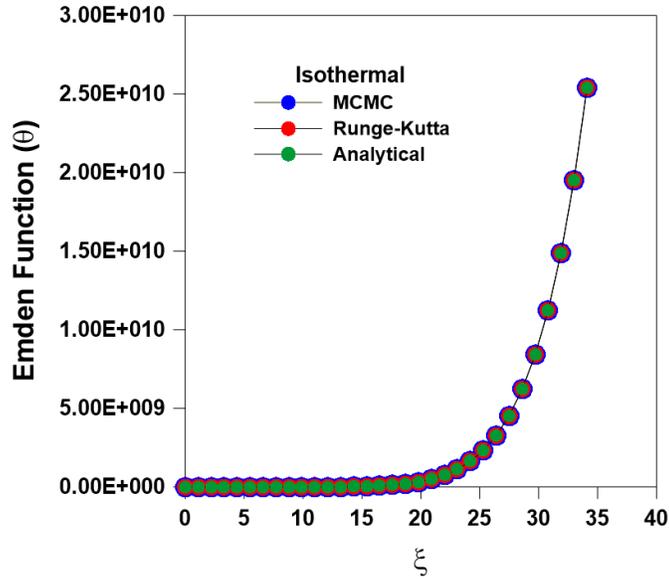

Figure 4: Comparison between the MC and numerical solutions computed for the Isothermal gas sphere.

## 4.4 White Dwarf equation

We solved Equation (22) for the range of the constant $C = 0.01 - 0.8$. The numerical results are presented in Tables A21-A25. The comparison with the numerical solutions is plotted in Figure 5, where we obtained an excellent agreement with the RK solutions. Interesting features could be read from the figure; as the parameter $C$ goes higher than zero (for $C = 0$, this case is analog to LE with $n = 3$), dimensionless distance $\xi$ and the Emden function $\theta(\xi)$ have values lesser than that of the polytrope with $n = 3$. Another important notice is that the white dwarf equation has not zeroth as that calculated for the polytropes with positive indices.

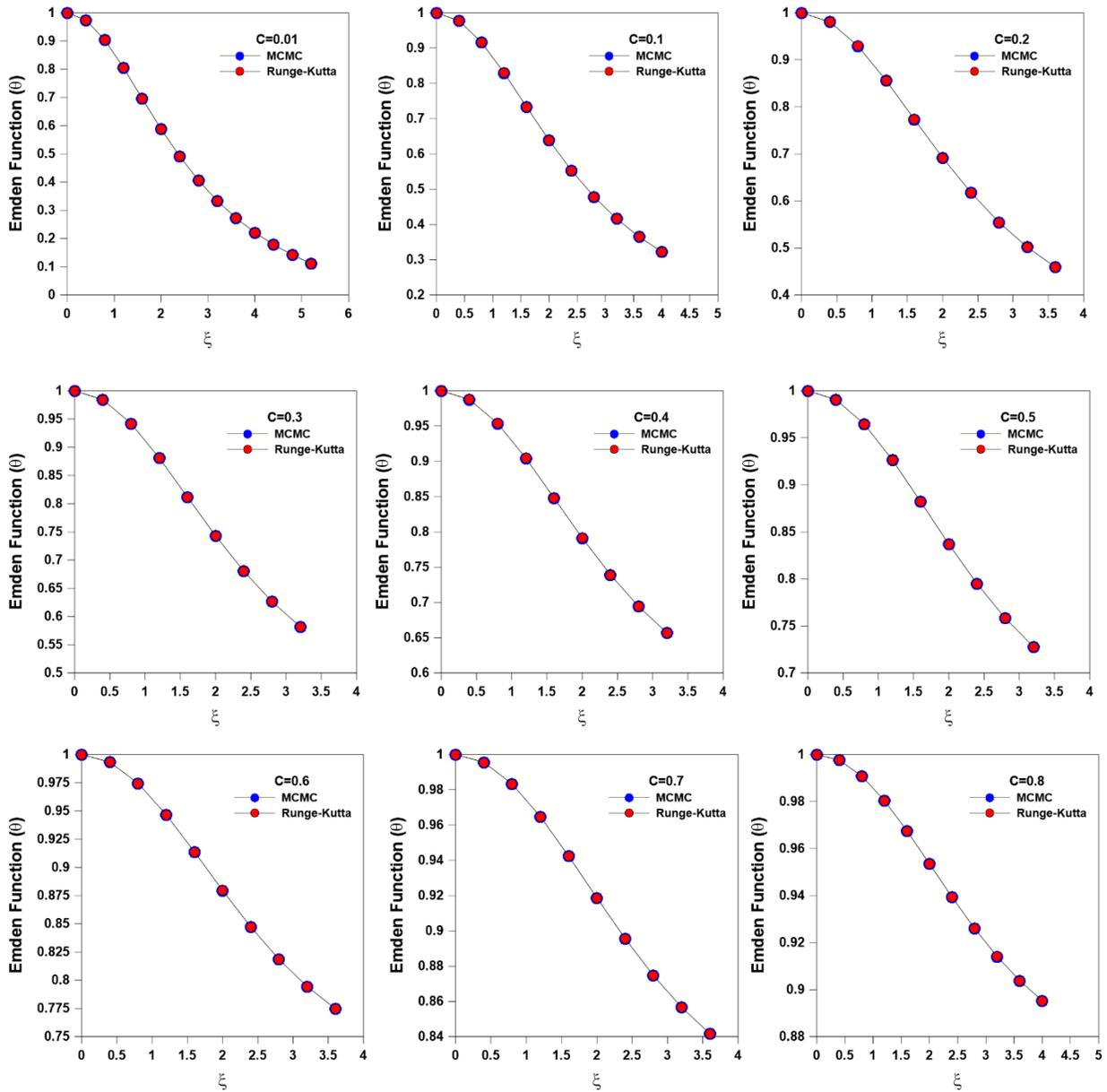

Figure 5: Comparison between the MC and numerical solutions computed for the White dwarf equations.

## 4.5 Comparative Study

To declare the impact of the proposed MC algorithm on the accuracy of the computed Emden function (*u*), we calculated the absolute errors between the MC solution and some numerical methods. Those methods are ANN based on an active-set algorithm (AST-NN, Ahmad et al. 2017), Chebyshev neural network (Ch-NN, Mall and Chakraverty 2014), Pattern search optimization technique (PS, Lewis et al. 2000), and Genetic algorithms (GA, Ahmad et al. 2016). We limited the comparison for polytropes with exact solutions only (i.e.polytropes with indexes n=0, n=1, and n=5; Equations (11)). Comparisons between these solutions are represented in Tables (2-4) and graphical representations for the absolute errors are shown in Figures (6-8) for the range of the dimensional parameter $\xi = 0 - 1$. For the polytropic index n=0, the comparison revealed that the maximum absolute error between the MC and the exact solution (labeled on the figure as MC-E) is 0.00033, 0.000014 for the AST-NN algorithm (AST-E), 0.0054 for the Ch-NN algorithm (Ch-E), 0.00056 for the PS algorithm (PS-E), and 0.00021 for the Genetic algorithm (GA-E). The comparison for n=1 showed that the maximum absolute error between the MC and the exact solution is 0.00029, 0.0071 for the AST-NN algorithm, 0.0041 for the Ch-NN algorithm, 0.0068 for the PS algorithm, and 0.0072 for the Genetic algorithm. Finally, the comparison for n=5 revealed that the maximum absolute error between the MC and the exact solution is 0.00019, 0.000012 for the AST-NN algorithm, 0.022 for the Ch-NN algorithm, 0.017 for the PS algorithm, and 0.017 for the Genetic algorithm.

Also, we compared the MC solution for the polytropic index with n=3 with the results from different methods. Table 5 shows a comparison of the numerical results applying MC, Homotopy Analysis Method (HAM), Padé approximants (PA) of an order [4; 4], and the numerical solution with the Simpson rule (SIMP); Al-Hayani et al. (2017). Figure 9 plots the absolute errors between the MC values and the methods mentioned above. The figure shows that the maximum absolute error between the three and the MC methods is about 0.00025.

Table 2: Comparison between different numerical methods and the exact method for n=0.

| $\xi$ | Exact | MC | AST-NN | Ch-NN | PS | GA |
|---|---|---|---|---|---|---|
| 0 | 1 | 1 | 1.000000 | 1.0000 | 1.000001 | 1.000037 |
| 0.1 | 0.998333 | 0.998299 | 0.998337 | 0.9993 | 0.998322 | 0.998451 |
| 0.2 | 0.993333 | 0.993266 | 0.993344 | 0.9901 | 0.993513 | 0.993512 |
| 0.3 | 0.98500 | 0.984900 | 0.985014 | 0.9822 | 0.985395 | 0.985151 |
| 0.4 | 0.973333 | 0.973200 | 0.973347 | 0.9766 | 0.973868 | 0.973433 |
| 0.5 | 0.958333 | 0.958166 | 0.958344 | 0.9602 | 0.958901 | 0.958420 |
| 0.6 | 0.94000 | 0.939800 | 0.940009 | 0.9454 | 0.940514 | 0.940122 |
| 0.7 | 0.918333 | 0.918100 | 0.918343 | 0.9134 | 0.918761 | 0.918511 |
| 0.8 | 0.893333 | 0.893066 | 0.893345 | 0.8892 | 0.893700 | 0.893543 |
| 0.9 | 0.865000 | 0.864700 | 0.865012 | 0.8633 | 0.865370 | 0.865192 |
| 1.0 | 0.833333 | 0.833000 | 0.833344 | 0.8322 | 0.833750 | 0.833480 |

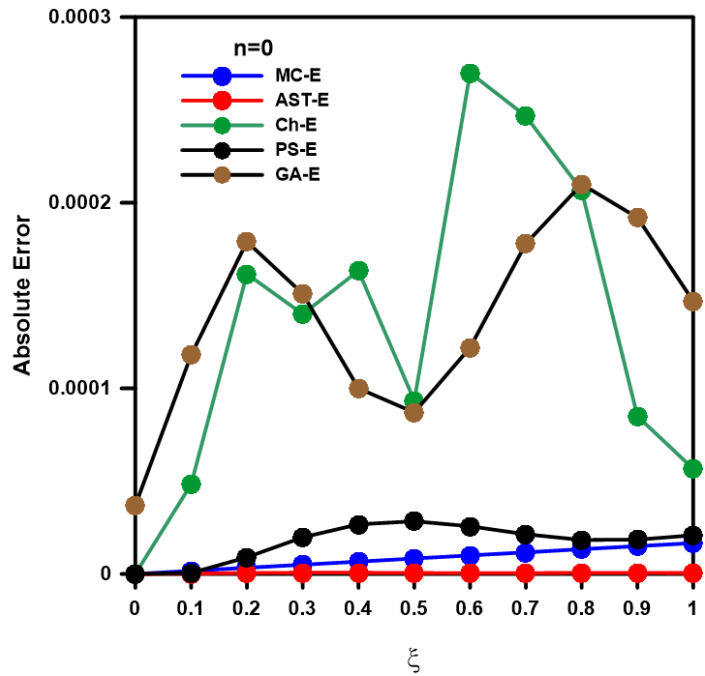

Figure 6: Comparative study based on values of absolute errors for polytrope with n=0.

Table 3: Comparison between different numerical methods and the exact method for n=1.

| $\xi$ | Exact | MC | AST-NN | Ch-NN | PS | GA |
|---|---|---|---|---|---|---|
| 0 | 1 | 1 | 1.000000 | 1 | 0.99998 | 0.99999 |
| 0.1 | 0.998334 | 0.998300 | 0.998337 | 1.00180 | 0.99805 | 0.99828 |
| 0.2 | 0.993346 | 0.993280 | 0.993364 | 0.99050 | 0.99278 | 0.99346 |
| 0.3 | 0.985067 | 0.984968 | 0.985138 | 0.98390 | 0.98470 | 0.98542 |
| 0.4 | 0.973545 | 0.973415 | 0.973757 | 0.97340 | 0.97343 | 0.97418 |
| 0.5 | 0.958851 | 0.958689 | 0.959355 | 0.95980 | 0.95891 | 0.95983 |
| 0.6 | 0.941070 | 0.940879 | 0.942096 | 0.94170 | 0.94169 | 0.94255 |
| 0.7 | 0.920310 | 0.920090 | 0.922172 | 0.92100 | 0.92187 | 0.92256 |
| 0.8 | 0.896695 | 0.896447 | 0.899800 | 0.89250 | 0.89943 | 0.90019 |
| 0.9 | 0.870363 | 0.870090 | 0.875212 | 0.87000 | 0.87489 | 0.87544 |
| 1.0 | 0.841470 | 0.841175 | 0.848656 | 0.84300 | 0.84832 | 0.84868 |

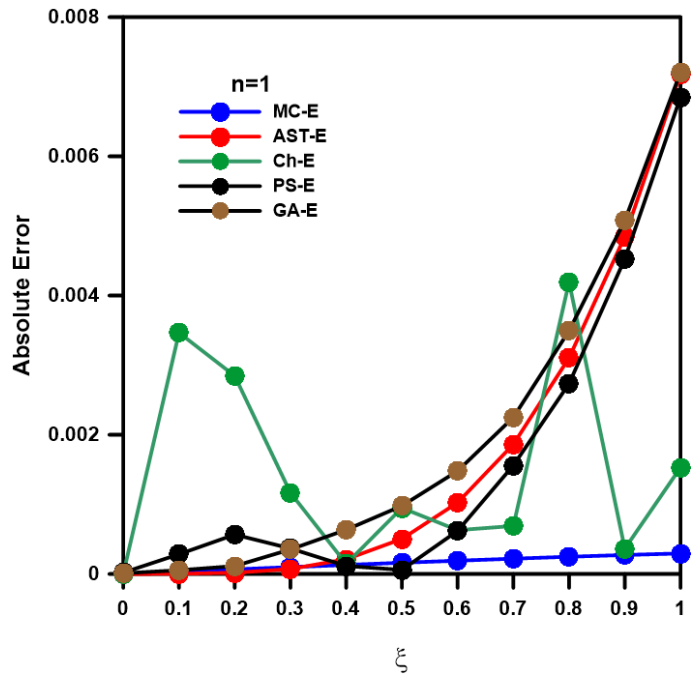

Figure 7: Comparative study based on values of absolute errors for polytrope with n=1.

Table 4: Comparison between different numerical methods and the exact method for n=5.

| $\xi$ | Exact | MC | AST-NN | Ch-NN | PS | GA |
|---|---|---|---|---|---|---|
| 0 | 1 | 1 | 1.000000 | 1.00000 | 0.99998 | 0.99999 |
| 0.1 | 0.998337 | 0.998304 | 0.998342 | 1.00180 | 0.99805 | 0.99828 |
| 0.2 | 0.993399 | 0.993334 | 0.993409 | 0.99050 | 0.99278 | 0.99346 |
| 0.3 | 0.985329 | 0.985234 | 0.985341 | 0.98390 | 0.98470 | 0.98542 |
| 0.4 | 0.974354 | 0.974233 | 0.974364 | 0.97340 | 0.97343 | 0.97418 |
| 0.5 | 0.9607689 | 0.960624 | 0.960776 | 0.95980 | 0.95891 | 0.95983 |
| 0.6 | 0.9449111 | 0.944747 | 0.944917 | 0.94170 | 0.94169 | 0.94255 |
| 0.7 | 0.9271455 | 0.926967 | 0.927151 | 0.92100 | 0.92187 | 0.92256 |
| 0.8 | 0.9078412 | 0.907652 | 0.907847 | 0.89250 | 0.89943 | 0.90019 |
| 0.9 | 0.8873565 | 0.887161 | 0.887362 | 0.87000 | 0.87489 | 0.87544 |
| 1.0 | 0.8660254 | 0.865827 | 0.866030 | 0.8431 | 0.84832 | 0.84868 |

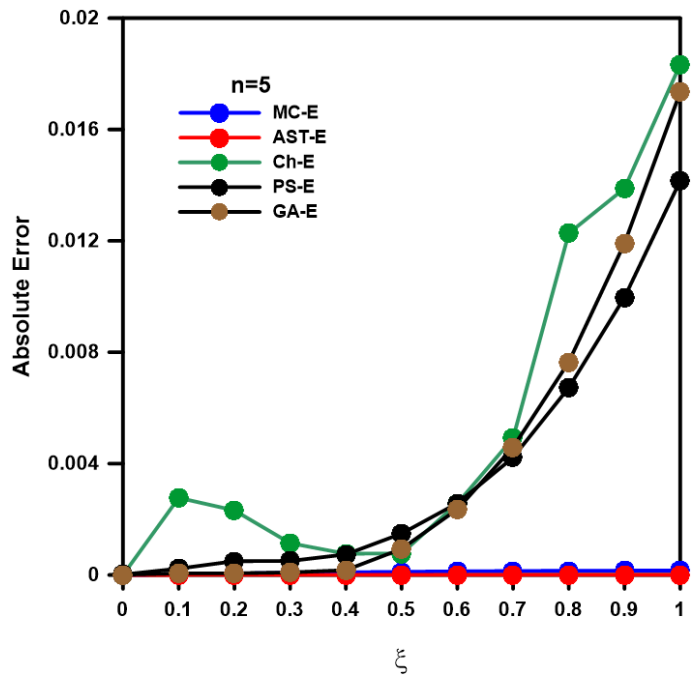

Figure 8: Comparative study based on values of absolute errors for polytrope with n=5.

Table 5: Comparison between different numerical methods and the exact method for n=3.

| $\xi$ | MC | HAM | PA | SIMP |
|---|---|---|---|---|
| 0 | 1 | 1 | 1 | 1 |
| 0.1 | 0.998369 | 0.998335 | 0.998335 | 0.998335 |
| 0.2 | 0.993438 | 0.993373 | 0.993373 | 0.993373 |
| 0.3 | 0.985295 | 0.985199 | 0.985199 | 0.985199 |
| 0.4 | 0.974081 | 0.973958 | 0.973958 | 0.973958 |
| 0.5 | 0.959839 | 0.959839 | 0.959839 | 0.959839 |
| 0.6 | 0.943240 | 0.943073 | 0.943073 | 0.943073 |
| 0.7 | 0.924106 | 0.923925 | 0.923922 | 0.923924 |
| 0.8 | 0.902867 | 0.902680 | 0.902672 | 0.902679 |
| 0.9 | 0.879819 | 0.879643 | 0.879617 | 0.879641 |
| 1.0 | 0.855057 | 0.855132 | 0.855057 | 0.855125 |

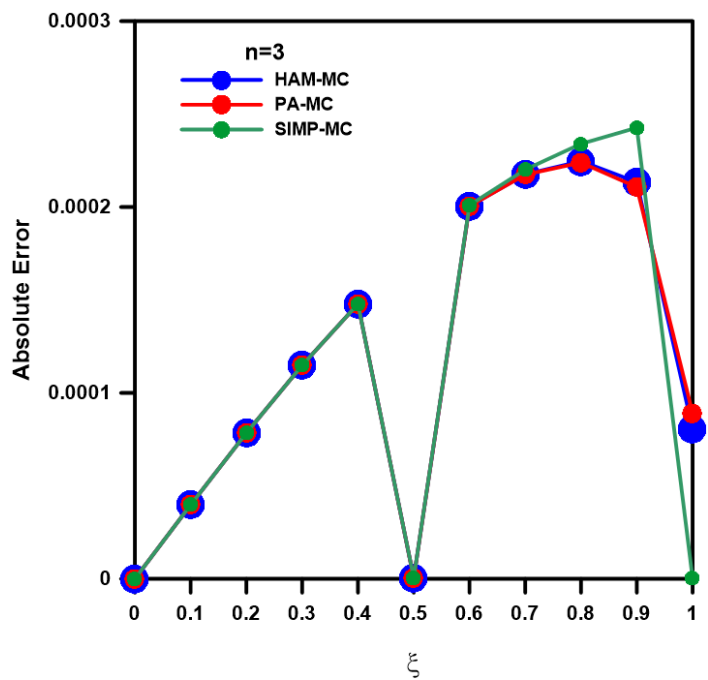

Figure 9: Comparative study based on values of absolute errors for polytrope with n=3.

## 5. Conclusion

The present paper introduces an MC solver for LE type equations in astrophysics, physics, and chemistry. We calculated eleven (i.e., eleven LE equations) models of the positive index polytropes (for polytropic indices $n = 0, 0.5, 1, 1.5, 2, 2.5, 3, 3.5, 4, 4.5, 4.99, 5$), nine models for the negative index polytrope ($n = -0.95, -0.99, -1.01, -1.5, -2, -3, -4, -5, -10$), the isothermal gas sphere, and nine models for the white dwarf equation. A total of $1000000$ random samples are utilized for the MC calculations. The Runge-Kutta approach (RK) for solving ordinary differential equations is used to numerically solve the four differential equations using the R package function rkMethod. We use an oddly spaced spacing with 0.001 increments.

Comparing the Emden function calculated by the MC, the RK, and the APS findings give good agreement for the polytropic indices $n = -0.5, 1.5, 2, 2.5 \& 3$, respectively. The zeroth of the Emden function obtained using the MC algorithm has been compared with that of the exact and the RK solutions. Relative errors revealed good agreement between the two solutions with a maximum value of 0.04%. The calculation for the negative index polytropes gives two interesting features; for $n = -0.95$ and $n = -0.99$, we obtained Emden functions to behave as the positive index polytropes, while for $n < -1$, the Emden functions act as monotonic functions. For the polytropic indexes n=0,1,3, and 5, we compared the MC results with several numerical approaches and found a good agreement.

**Acknowledgments**

This paper is based upon work supported by Science, Technology & Innovation Funding Authority (STDF) under Grant number 37122. We are so grateful to the reviewers for many valuable suggestions and comments that significantly improved the paper.

**Appendix A: Numerical Results**

We list in the following tables the numerical results obtained for the four differential equations, Equations (10), (14), and (21), for the positive index polytrope, negative index polytrope, the isothermal gas sphere, and the white dwarf equation, respectively. We listed tables for some selected values; complete tables may be requested from the authors. The designation of the columns is as follows:

Column 1: The dimensionless distance ($\xi$).

Column 2: The Emden function calculated by the exact solution ($\theta_{Exact}$) or RK solution ($\theta_{RK}$).

Column 3: The Emden function calculated by the MC solution ($\theta_{MC}$).

Column 4: The relative error computed for the MC and RK solutions.

Table A1: Comparison between Exact and MC solutions of
$\theta(\xi)$ with polytrope index $n = 0$.

| $\xi$ | $\theta_{Exact}$ | $\theta_{MC}$ | Relative error |
|---|---|---|---|
| 0.000 | 1.00000e+00 | 1.00000e+00 | 0.000000e+00 |
| 0.300 | 0.9850000 | 0.9849000 | 1.014802e-04 |
| 0.600 | 0.9400000 | 0.9398001 | 0.0002126788 |
| 0.900 | 0.8650000 | 0.8647001 | 0.0003466991 |
| 1.200 | 0.7600000 | 0.7596001 | 0.0005261556 |
| 1.500 | 0.6250000 | 0.6245001 | 0.0007997844 |
| 1.800 | 0.4600000 | 0.4594001 | 0.0013040314 |
| 2.100 | 0.2650000 | 0.2643002 | 0.002640928 |
| 2.400 | 0.0400000000 | 0.0392001615 | 0.019995961 |
| 2.440 | 0.0077333333 | 0.0069201625 | 0.105151406 |

Table A2: Comparison between Exact and MC solutions of
$\theta(\xi)$ with polytrope index $n = 1$.

| $\xi$ | $\theta_{Exact}$ | $\theta_{MC}$ | Relative error |
|---|---|---|---|
| 0.000 | 1.00000e+00 | 1.00000e+00 | 0.000000e+00 |
| 0.400 | 0.9735459 | 0.9734151 | 1.343466e-04 |
| 0.800 | 0.8966951 | 0.8964480 | 0.0002756068 |
| 1.200 | 0.7766992 | 0.7763629 | 0.0004330656 |
| 1.600 | 0.6247335 | 0.6243443 | 0.0006229333 |
| 2.000 | 0.4546487 | 0.4542480 | 0.0008814356 |
| 2.400 | 0.2814430 | 0.2810717 | 0.001319296 |
| 2.800 | 0.11963863 | 0.11933265 | 0.002557471 |
| 3.120 | 0.0069201845 | 6.686209e-03 | 0.033810660 |
| 3.130 | 0.0037036402 | 3.472125e-03 | 0.062510190 |
| 3.140 | 0.0005072143 | 2.781704e-04 | 0.451572344 |

Table A3: Comparison between Exact and MC solutions of $\theta(\xi)$ with polytrope index $n = 5$.

| $\xi$ | $\theta_{Exact}$ | $\theta_{MC}$ | Relative error | $\xi$ | $\theta_{Exact}$ | $\theta_{MC}$ | Relative error |
|---|---|---|---|---|---|---|---|
| 0.000 | 1.00000e+00 | 1.00000e+00 | 0.000000e+00 | 30.00 | 0.05763904 | 0.05778058 | 0.002455610 |
| 1.000 | 0.8660254 | 0.8658271 | 0.0002289966 | 35.00 | 0.04942668 | 0.04956891 | 0.002877580 |
| 2.000 | 0.6546537 | 0.6545291 | 0.0001903568 | 40.00 | 0.04326073 | 0.04340341 | 0.003298076 |
| 3.000 | 0.5000000 | 0.4999688 | 6.249792e-05 | 45.00 | 0.03846154 | 0.03860452 | 0.003717582 |
| 4.000 | 0.3973597 | 0.3973875 | 7.006908e-05 | 50.00 | 0.03462025 | 0.03476345 | 0.004136394 |
| 5.000 | 0.3273268 | 0.3273901 | 0.0001932484 | 55.00 | 0.03147623 | 0.03161959 | 0.004554699 |
| 6.000 | 0.2773501 | 0.2774354 | 0.0003073958 | 60.00 | 0.02885549 | 0.02899898 | 0.004972623 |
| 11.00 | 0.1555428 | 0.1556680 | 0.0008050144 | 65.00 | 0.02663748 | 0.02678106 | 0.005390254 |
| 15.00 | 0.114707867 | 0.114841684 | 0.001166591 | 70.00 | 0.02473601 | 0.02487967 | 0.005807653 |
| 20.00 | 0.08627960 | 0.08641790 | 0.001602920 | 75.00 | 0.02308785 | 0.02323157 | 0.006224867 |
| 22.00 | 0.07848671 | 0.07862602 | 0.001774929 | 80.00 | 0.02164556 | 0.02178933 | 0.006641930 |
| 24.00 | 0.07198158 | 0.07212165 | 0.001946023 | 85.00 | 0.02037284 | 0.02051665 | 0.007058869 |
| 26.00 | 0.06647001 | 0.06661069 | 0.002116408 | 90.00 | 0.01924145 | 0.01938529 | 0.007475704 |
| 28.00 | 0.06174094 | 0.06188210 | 0.002286234 | 95.00 | 0.01822908 | 0.01837296 | 0.007892451 |
| 29.00 | 0.059619648 | 0.059761004 | 0.002370972 | 100.0 | 0.01731791 | 0.01746181 | 0.008309124 |

Table A4: Comparison between RK and MC solutions of

$\theta(\xi)$ with polytrope index $n = 0.5$

| $\xi$ | $\theta_{RK}$ | $\theta_{MC}$ | Relative error |
|---|---|---|---|
| 0.000 | 1.00000e+00 | 1.00000e+00 | 0.000000e+00 |
| 0.400 | 0.9734406 | 0.9734394 | 1.221633e-06 |
| 0.800 | 0.8950494 | 0.8950433 | 6.816313e-06 |
| 1.200 | 0.7687456 | 0.7687261 | 2.537468e-05 |
| 1.600 | 0.6012647 | 0.6012191 | 7.583467e-05 |
| 2.000 | 0.4025799 | 0.4024912 | 0.0002201448 |
| 2.400 | 0.1869203 | 0.1867668 | 0.0008210544 |
| 2.740 | 0.006375151 | 0.006141292 | 0.036682865 |
| 2.750 | 0.001350444 | 0.001113427 | 0.175510805 |

Table A5: Comparison between RK and MC solutions of

$\theta(\xi)$ with polytrope index $n = 1.5$.

| $\xi$ | $\theta_{RK}$ | $\theta_{MC}$ | Relative error |
|---|---|---|---|
| 0.000 | 1.00000e+00 | 1.00000e+00 | 0.000000e+00 |
| 0.500 | 0.9591043 | 0.9592605 | 0.0001628245 |
| 1.000 | 0.8451701 | 0.8454285 | 0.0003056639 |
| 1.500 | 0.6811246 | 0.6814026 | 0.0004081928 |
| 2.000 | 0.4959369 | 0.4961598 | 0.0004494538 |
| 2.500 | 0.3158926 | 0.3160176 | 0.0003957444 |
| 3.000 | 0.1588576 | 0.1588785 | 1.315799e-04 |
| 3.500 | 0.032615649 | 0.032552086 | 0.0019488652 |
| 3.600 | 1.109091e-02 | 1.101400e-02 | 0.006934564 |
| 3.652 | 3.566154e-04 | 2.732765e-04 | 0.233694247 |

Table A6: Comparison between RK and MC solutions of

$\theta(\xi)$ with polytrope index $n = 2$.

| $\xi$ | $\theta_{RK}$ | $\theta_{MC}$ | Relative error |
|---|---|---|---|
| 0.000 | 1.00000e+00 | 1.00000e+00 | 0.000000e+00 |
| 0.500 | 0.9593532 | 0.9595062 | 0.0001594812 |
| 1.000 | 0.8486545 | 0.8488931 | 0.0002811817 |
| 1.500 | 0.6953674 | 0.6956018 | 0.0003371824 |
| 2.000 | 0.5298365 | 0.5300028 | 0.0003138883 |
| 2.500 | 0.3747393 | 0.3748155 | 0.0002034021 |
| 3.000 | 0.2418240 | 0.2418195 | 1.887474e-05 |
| 3.500 | 0.1339689 | 0.1339059 | 0.0004699780 |
| 4.000 | 0.04884002 | 0.04874116 | 0.002024153 |
| 4.350 | 3.658919e-04 | 2.523544e-04 | 0.310303453 |

Table A7: Comparison between RK and MC solutions of
$\theta(\xi)$ with polytrope index $n = 2.5$.

| $\xi$ | $\theta_{RK}$ | $\theta_{MC}$ | Relative error |
|---|---|---|---|
| 0.000 | 1.00000e+00 | 1.00000e+00 | 0.000000e+00 |
| 0.500 | 0.9595982 | 0.9597481 | 1.562349e-04 |
| 1.000 | 0.8519445 | 0.8521654 | 0.0002593079 |
| 1.500 | 0.7080725 | 0.7082719 | 0.0002816049 |
| 2.000 | 0.5583724 | 0.5584983 | 0.0002254515 |
| 2.000 | 0.5583724 | 0.5584983 | 0.0002254515 |
| 2.500 | 0.4220076 | 0.4220525 | 1.064585e-04 |
| 3.000 | 0.3066750 | 0.3066553 | 6.419570e-05 |
| 3.500 | 0.2128325 | 0.2127699 | 0.0002938066 |
| 4.000 | 0.1376806 | 0.1375936 | 0.0006316995 |
| 4.500 | 0.07755395 | 0.07745543 | 0.001270399 |
| 5.000 | 0.02901904 | 0.02891703 | 0.003515047 |
| 5.350 | 4.025760e-04 | 3.010598e-04 | 0.252166581 |

Table A8: Comparison between RK and MC solutions of
$\theta(\xi)$ with polytrope index $n = 3$.

| $\xi$ | $\theta_{RK}$ | $\theta_{MC}$ | Relative error |
|---|---|---|---|
| 0.000 | 1.0000e+00 | 1.0000e+00 | 0.00000e+00 |
| 0.500 | 0.9598395 | 0.9599864 | 0.0001530715 |
| 1.000 | 0.8550579 | 0.8552628 | 0.0002396530 |
| 1.500 | 0.7195020 | 0.7196727 | 0.0002372486 |
| 2.000 | 0.5828505 | 0.5829465 | 0.0001646699 |
| 2.500 | 0.4611265 | 0.4611503 | 5.164349e-05 |
| 3.000 | 0.3592264 | 0.3591976 | 8.023693e-05 |
| 3.500 | 0.2762625 | 0.2762013 | 0.0002215195 |
| 4.000 | 0.2092815 | 0.2092034 | 0.0003732390 |
| 4.500 | 0.1550692 | 0.1549845 | 0.0005463745 |
| 5.000 | 0.1108197 | 0.1107347 | 0.0007669683 |
| 5.500 | 0.07428603 | 0.07420445 | 0.001098213 |
| 6.000 | 0.04373784 | 0.04366168 | 0.001741298 |
| 6.500 | 0.01786603 | 0.01779629 | 0.003903565 |
| 6.870 | 1.143486e-03 | 1.078780e-03 | 0.056586346 |
| 6.880 | 7.164872e-04 | 6.519191e-04 | 0.090117514 |
| 6.890 | 2.907277e-04 | 2.262974e-04 | 0.221617445 |

Table A9: Comparison between RK and MC solutions of
$\theta(\xi)$ with polytrope index $n = 3.5$.

| $\xi$ | $\theta_{RK}$ | $\theta_{MC}$ | Relative error |
|---|---|---|---|
| 0.000 | 1.00000e+00 | 1.00000e+00 | 0.000000e+00 |
| 0.500 | 0.9600772 | 0.9599264 | 0.0001570141 |
| 1.000 | 0.8580099 | 0.8577821 | 0.0002654604 |
| 2.000 | 0.6041619 | 0.6039827 | 0.0002966481 |
| 3.000 | 0.4029444 | 0.4028763 | 0.0001688633 |
| 4.000 | 0.2683486 | 0.2683612 | 4.671913e-05 |
| 5.000 | 0.1786841 | 0.1787481 | 0.0003581215 |
| 6.000 | 0.1166472 | 0.1167453 | 0.0008401983 |
| 7.000 | 0.07180062 | 0.07192273 | 0.001700682 |

| ξ | | | |
|---|---|---|---|
| 8.000 | 0.03805984 | 0.03820017 | 0.003687073 |
| 9.000 | 0.011802996 | 0.011957768 | 0.013112958 |
| 9.500 | 7.471099e-04 | 0.0009080641 | 0.21543566 |
| 9.510 | 5.378502e-04 | 0.0006989220 | 0.29947324 |
| 9.520 | 3.290301e-04 | 0.0004902193 | 0.48989177 |
| 9.530 | 1.206483e-04 | 0.0002819546 | 1.33699652 |

Table A10: Comparison between RK and MC solutions of

$\theta(\xi)$ with polytrope index $n = 4$.

| $\xi$ | $\theta_{RK}$ | $\theta_{MC}$ | Relative error |
|---|---|---|---|
| 0.000 | 1.00000e+00 | 1.00000e+00 | 0.000000e+00 |
| 1.000 | 0.8608141 | 0.8605968 | 0.0002523413 |
| 2.000 | 0.6229408 | 0.6227829 | 0.0002533719 |
| 3.000 | 0.4400506 | 0.4399975 | 0.0001205621 |
| 4.000 | 0.3180423 | 0.3180612 | 5.952305e-05 |
| 5.000 | 0.2359226 | 0.2359864 | 0.0002706464 |
| 6.000 | 0.1783841 | 0.1784770 | 0.0005207540 |
| 7.000 | 0.1363522 | 0.1364650 | 0.0008277002 |
| 8.000 | 0.1045040 | 0.1046314 | 0.001219358 |
| 9.000 | 0.07961936 | 0.07975797 | 0.001740897 |
| 10.000 | 0.05967264 | 0.05982018 | 0.002472474 |
| 11.000 | 0.04333999 | 0.04349488 | 0.003573840 |
| 12.000 | 0.02972583 | 0.02988692 | 0.005418869 |
| 13.000 | 0.01820531 | 0.01837169 | 0.009139184 |
| 14.000 | 0.008330434 | 0.008501414 | 0.02052474 |
| 14.600 | 0.003054810 | 0.003228268 | 0.05678208 |
| 14.930 | 3.339586e-04 | 0.0005087013 | 0.5232468 |
| 14.940 | 2.533849e-04 | 0.0004281658 | 0.6897839 |
| 14.950 | 1.729190e-04 | 0.0003477379 | 1.0109869 |
| 14.960 | 9.256075e-05 | 0.0002674176 | 1.8891043 |

Table A11: Comparison between RK and MC solutions of
$\theta(\xi)$ with polytrope index $n = 4.5$.

| $\xi$ | $\theta_{RK}$ | $\theta_{MC}$ | Relative error |
|---|---|---|---|
| 0.000 | 1.00000e+00 | 1.00000e+00 | 0.000000e+00 |
| 2.000 | 0.6396537 | 0.6395138 | 0.0002186821 |
| 4.000 | 0.3605337 | 0.3605576 | 6.636828e-05 |
| 6.000 | 0.2313630 | 0.2314518 | 0.0003836721 |
| 8.000 | 0.1617328 | 0.1618502 | 0.0007261914 |
| 10.000 | 0.1189406 | 0.1190736 | 0.001117632 |
| 12.000 | 0.09015570 | 0.09029831 | 0.001581835 |
| 14.000 | 0.06952037 | 0.06966968 | 0.002147665 |
| 16.000 | 0.05402041 | 0.05417468 | 0.002855762 |
| 18.000 | 0.04195725 | 0.04211538 | 0.003768835 |
| 20.000 | 0.03230423 | 0.03246547 | 0.004991228 |
| 24.000 | 0.01782304 | 0.01798900 | 0.009311670 |
| 28.000 | 0.007479021 | 0.007648421 | 0.02265005 |
| 30.500 | 0.002391790 | 0.002562903 | 0.07154183 |

| 31.000 | 0.001472806 | 0.001644230 | 0.1163928 |
| 31.500 | 0.0005829964 | 0.0007547220 | 0.2945569 |
| 31.600 | 0.0004084135 | 0.0005801983 | 0.4206150 |
| 31.700 | 2.349320e-04 | 0.0004067757 | 0.7314613 |
| 31.800 | 6.254169e-05 | 0.0002344439 | 2.7486018 |

Table A12: Comparison between RK and MC solutions of $\theta(\xi)$ with polytrope index $n = 4.99$.

| $\xi$ | $\theta_{RK}$ | $\theta_{MC}$ | Relative error |
|---|---|---|---|
| 0.000 | 1.00000e+00 | 1.00000e+00 | 0.000000e+00 |
| 2.000 | 0.6543685 | 0.6542436 | 0.0001908621 |
| 4.000 | 0.3966708 | 0.3966986 | 7.005805e-05 |
| 6.000 | 0.2764896 | 0.2765749 | 0.0003086424 |
| 8.000 | 0.2106689 | 0.2107784 | 0.0005199808 |
| 10.000 | 0.1696935 | 0.1698151 | 0.0007166558 |
| 12.000 | 0.1418680 | 0.1419964 | 0.0009052283 |
| 14.000 | 0.1217825 | 0.1219151 | 0.001089043 |
| 16.000 | 0.1066194 | 0.1067548 | 0.001269925 |
| 18.000 | 0.09477443 | 0.09491176 | 0.001448942 |
| 20.000 | 0.08526971 | 0.08540843 | 0.001626755 |
| 22.000 | 0.07747604 | 0.07761579 | 0.001803793 |
| 24.000 | 0.07097066 | 0.07111121 | 0.001980347 |
| 26.000 | 0.06545920 | 0.06560037 | 0.002156621 |
| 28.000 | 0.06073046 | 0.06087213 | 0.002332762 |
| 30.000 | 0.05662902 | 0.05677110 | 0.002508878 |
| 35.000 | 0.04841810 | 0.04856091 | 0.002949552 |
| 40.000 | 0.04225369 | 0.04239699 | 0.003391392 |
| 45.000 | 0.03745597 | 0.03759961 | 0.003834885 |
| 50.000 | 0.03361604 | 0.03375993 | 0.004280331 |
| 55.000 | 0.03047326 | 0.03061734 | 0.004727931 |
| 60.000 | 0.02785364 | 0.02799786 | 0.005177824 |
| 65.000 | 0.02563663 | 0.02578097 | 0.005630113 |
| 70.000 | 0.02373607 | 0.02388050 | 0.006084878 |
| 75.000 | 0.02208874 | 0.02223324 | 0.006542185 |
| 80.000 | 0.02064719 | 0.02079176 | 0.007002088 |
| 85.000 | 0.01937514 | 0.01951977 | 0.007464636 |
| 90.000 | 0.01824436 | 0.01838904 | 0.007929870 |
| 95.000 | 0.01723256 | 0.01737728 | 0.008397830 |
| 100.00 | 0.01632191 | 0.01646666 | 0.008868552 |

Table A13: Comparison between RK and MC solutions of $\theta(\xi)$ with polytrope index $n = -1.01$.

| $\xi$ | $\theta_{RK}$ | $\theta_{MC}$ | Relative error | $\xi$ | $\theta_{RK}$ | $\theta_{MC}$ | Relative error |
|---|---|---|---|---|---|---|---|
| 0.000 | 1.00000e+00 | 1.00000e+00 | 0.000000e+00 | 23.000 | 16.06488 | 16.06519 | 1.907790e-05 |
| 1.000 | 1.159022 | 1.159321 | 0.0002581444 | 24.000 | 16.76643 | 16.76673 | 1.805431e-05 |
| 2.000 | 1.569078 | 1.569547 | 0.0002987201 | 25.000 | 17.46751 | 17.46781 | 1.713843e-05 |
| 3.000 | 2.121349 | 2.121882 | 0.0002513584 | 26.000 | 18.16817 | 18.16846 | 1.631522e-05 |
| 4.000 | 2.746491 | 2.747036 | 0.0001985520 | 27.000 | 18.86841 | 18.8687 | 1.557216e-05 |
| 5.000 | 3.409439 | 3.409973 | 0.0001568384 | 28.000 | 19.56826 | 19.56855 | 1.489877e-05 |
| 6.000 | 4.092713 | 4.093228 | 0.0001258860 | 29.000 | 20.26774 | 20.26803 | 1.428621e-05 |
| 7.000 | 4.787235 | 4.787728 | 0.0001029766 | 30.000 | 20.96687 | 20.96716 | 1.372702e-05 |
| 8.000 | 5.488061 | 5.488532 | 8.578283e-05 | 31.000 | 21.66566 | 21.66595 | 1.321484e-05 |
| 9.000 | 6.192386 | 6.192836 | 7.264446e-05 | 32.000 | 22.36415 | 22.36443 | 1.274422e-05 |
| 10.000 | 6.898565 | 6.898996 | 6.242207e-05 | 33.000 | 23.06233 | 23.06261 | 1.231049e-05 |
| 11.000 | 7.605613 | 7.606026 | 5.433329e-05 | 34.000 | 23.76023 | 23.76051 | 1.190962e-05 |
| 12.000 | 8.312928 | 8.313325 | 4.783399e-05 | 35.000 | 24.45785 | 24.45814 | 1.153811e-05 |
| 13.000 | 9.020142 | 9.020526 | 4.253929e-05 | 36.000 | 25.15522 | 25.15551 | 1.119292e-05 |
| 14.000 | 9.727034 | 9.727405 | 3.817210e-05 | 37.000 | 25.85235 | 25.85263 | 1.087141e-05 |
| 15.000 | 10.43347 | 10.43383 | 3.452948e-05 | 38.000 | 26.54924 | 26.54952 | 1.057124e-05 |
| 16.000 | 11.13938 | 11.13973 | 3.146041e-05 | 39.000 | 27.24592 | 27.24620 | 1.029038e-05 |
| 17.000 | 11.84471 | 11.84506 | 2.885078e-05 | 40.000 | 27.94238 | 27.94266 | 1.002704e-05 |
| 18.000 | 12.54947 | 12.54980 | 2.661324e-05 | 42.000 | 29.33470 | 29.33498 | 9.546672e-06 |
| 19.000 | 13.25365 | 13.25398 | 2.468003e-05 | 46.000 | 32.11719 | 32.11747 | 8.736822e-06 |
| 20.000 | 13.95726 | 13.95758 | 2.299800e-05 | 48.000 | 33.50747 | 33.50775 | 8.391986e-06 |
| 21.000 | 14.66032 | 14.66064 | 2.152504e-05 | 50.000 | 34.89717 | 34.89746 | 8.079390e-06 |
| 22.000 | 15.36285 | 15.36316 | | | | | |

Table A14: Comparison between RK and MC solutions of $\theta(\xi)$ with polytrope index $n = -1.5$.

| $\xi$ | $\theta_{RK}$ | $\theta_{MC}$ | Relative error | $\xi$ | $\theta_{RK}$ | $\theta_{MC}$ | Relative error |
|---|---|---|---|---|---|---|---|
| 0.000 | 1.00000e+00 | 1.00000e+00 | 0.000000e+00 | 23.000 | 10.62928 | 10.62936 | 7.758030e-06 |
| 1.000 | 1.155613 | 1.155898 | 0.0002463248 | 24.000 | 11.00046 | 11.00054 | 7.271363e-06 |
| 2.000 | 1.532601 | 1.533004 | 0.0002633452 | 25.000 | 11.36827 | 11.36834 | 6.852927e-06 |
| 3.000 | 2.003254 | 2.003669 | 0.0002071382 | 26.000 | 11.73286 | 11.73294 | 6.491663e-06 |
| 4.000 | 2.502164 | 2.502551 | 0.0001545821 | 27.000 | 12.09441 | 12.09449 | 6.178525e-06 |
| 5.000 | 3.003615 | 3.003963 | 0.0001159915 | 28.000 | 12.45307 | 12.45314 | 5.906069e-06 |
| 6.000 | 3.498098 | 3.498408 | 8.867979e-05 | 29.000 | 12.80895 | 12.80903 | 5.668138e-06 |
| 7.000 | 3.982396 | 3.982672 | 6.920466e-05 | 30.000 | 13.16221 | 13.16228 | 5.459613e-06 |
| 8.000 | 4.455784 | 4.456029 | 5.506055e-05 | 31.000 | 13.51293 | 13.51300 | 5.276216e-06 |
| 9.000 | 4.918525 | 4.918744 | 4.458176e-05 | 32.000 | 13.86124 | 13.86131 | 5.114363e-06 |
| 10.000 | 5.371255 | 5.371452 | 3.667168e-05 | 33.000 | 14.20723 | 14.20730 | 4.971033e-06 |
| 11.000 | 5.814722 | 5.814900 | 3.059915e-05 | 34.000 | 14.55099 | 14.55106 | 4.843671e-06 |
| 12.000 | 6.249670 | 6.249832 | 2.586713e-05 | 35.000 | 14.89261 | 14.89268 | 4.730111e-06 |
| 13.000 | 6.676797 | 6.676945 | 2.213073e-05 | 36.000 | 15.23217 | 15.23224 | 4.628509e-06 |
| 14.000 | 7.096737 | 7.096873 | 1.914580e-05 | 37.000 | 15.56974 | 15.56981 | 4.537290e-06 |
| 15.000 | 7.510061 | 7.510187 | 1.673635e-05 | 38.000 | 15.90540 | 15.90547 | 4.455106e-06 |
| 16.000 | 7.917277 | 7.917394 | 1.477332e-05 | 39.000 | 16.23920 | 16.23927 | 4.380800e-06 |
| 17.000 | 8.318837 | 8.318947 | 1.316066e-05 | 40.000 | 16.57121 | 16.57128 | 4.313375e-06 |
| 18.000 | 8.715146 | 8.715249 | 1.182582e-05 | 42.000 | 17.23008 | 17.23016 | 4.195845e-06 |
| 19.000 | 9.106563 | 9.106661 | 1.071339e-05 | 46.000 | 18.52869 | 18.52876 | 4.012420e-06 |
| 20.000 | 9.493410 | 9.493503 | 9.780505e-06 | 48.000 | 19.16914 | 19.16921 | 3.939115e-06 |
| 21.000 | 9.875977 | 9.876065 | 8.993683e-06 | 50.000 | 19.80410 | 19.80418 | 3.874571e-06 |
| 22.000 | 10.25452 | 10.25461 | | | | | |

Table A15: Comparison between RK and MC solutions of $\theta(\xi)$ with polytrope index $n = -2$.

| $\xi$ | $\theta_{RK}$ | $\theta_{MC}$ | Relative error | $\xi$ | $\theta_{RK}$ | $\theta_{MC}$ | Relative error |
|---|---|---|---|---|---|---|---|
| 0.000 | 1.00000e+00 | 1.00000e+00 | 0.000000e+00 | 23.000 | 7.836229 | 7.836236 | 8.970258e-07 |
| 1.000 | 1.152315 | 1.152586 | 0.0002351381 | 24.000 | 8.062923 | 8.062930 | 7.801629e-07 |
| 2.000 | 1.500526 | 1.500876 | 0.0002334171 | 25.000 | 8.286250 | 8.286256 | 7.033198e-07 |
| 3.000 | 1.907784 | 1.908113 | 0.0001724398 | 26.000 | 8.506403 | 8.506408 | 6.583673e-07 |
| 4.000 | 2.317285 | 2.317567 | 0.0001216847 | 27.000 | 8.723557 | 8.723563 | 6.387954e-07 |
| 5.000 | 2.712219 | 2.712453 | 8.639992e-05 | 28.000 | 8.937872 | 8.937878 | 6.393677e-07 |
| 6.000 | 3.088992 | 3.089185 | 6.235126e-05 | 29.000 | 9.149493 | 9.149499 | 6.558586e-07 |
| 7.000 | 3.448074 | 3.448232 | 4.573710e-05 | 30.000 | 9.358555 | 9.358561 | 6.848487e-07 |
| 8.000 | 3.791064 | 3.791193 | 3.402585e-05 | 31.000 | 9.565178 | 9.565185 | 7.235627e-07 |
| 9.000 | 4.119735 | 4.119840 | 2.560709e-05 | 32.000 | 9.769476 | 9.769484 | 7.697430e-07 |
| 10.000 | 4.435735 | 4.435821 | 1.944988e-05 | 33.000 | 9.971553 | 9.971562 | 8.215521e-07 |
| 11.000 | 4.740509 | 4.740580 | 1.488067e-05 | 34.000 | 10.17150 | 10.17151 | 8.774918e-07 |
| 12.000 | 5.035296 | 5.035354 | 1.144913e-05 | 35.000 | 10.36942 | 10.36943 | 9.363413e-07 |
| 13.000 | 5.321151 | 5.321199 | 8.847199e-06 | 36.000 | 10.56538 | 10.56539 | 9.971055e-07 |
| 14.000 | 5.598975 | 5.599013 | 6.859690e-06 | 37.000 | 10.75947 | 10.75948 | 1.058975e-06 |
| 15.000 | 5.869535 | 5.869566 | 5.333411e-06 | 38.000 | 10.95175 | 10.95176 | 1.121293e-06 |
| 16.000 | 6.133493 | 6.133519 | 4.157427e-06 | 39.000 | 11.14230 | 11.14231 | 1.183528e-06 |
| 17.000 | 6.391420 | 6.391441 | 3.250187e-06 | 40.000 | 11.33117 | 11.33118 | 1.245252e-06 |
| 18.000 | 6.643812 | 6.643829 | 2.550922e-06 | 42.000 | 11.70412 | 11.70414 | 1.365862e-06 |
| 19.000 | 6.891102 | 6.891116 | 2.013802e-06 | 46.000 | 12.43233 | 12.43235 | 1.589954e-06 |
| 20.000 | 7.133671 | 7.133682 | 1.603888e-06 | 48.000 | 12.78830 | 12.78833 | 1.691613e-06 |
| 21.000 | 7.371855 | 7.371864 | 1.294285e-06 | 50.000 | 13.13928 | 13.13931 | 1.785862e-06 |
| 22.000 | 7.605952 | 7.605960 | | | | | |

Table A16: Comparison between RK and MC solutions of $\theta(\xi)$ with polytrope index $n = -3$.

| $\xi$ | $\theta_{RK}$ | $\theta_{MC}$ | Relative error | $\xi$ | $\theta_{RK}$ | $\theta_{MC}$ | Relative error |
|---|---|---|---|---|---|---|---|
| 0.000 | 1.00000e+00 | 1.00000e+00 | 0.000000e+00 | 23.000 | 5.185568 | 5.185536 | 6.029996e-06 |
| 1.000 | 1.146205 | 1.146451 | 0.0002150764 | 24.000 | 5.296815 | 5.296785 | 5.697356e-06 |
| 2.000 | 1.448011 | 1.448282 | 0.0001869318 | 25.000 | 5.405624 | 5.405595 | 5.360659e-06 |
| 3.000 | 1.766015 | 1.766232 | 0.0001231185 | 26.000 | 5.512151 | 5.512123 | 5.024180e-06 |
| 4.000 | 2.061805 | 2.061964 | 7.740714e-05 | 27.000 | 5.616537 | 5.616511 | 4.691107e-06 |
| 5.000 | 2.331000 | 2.331112 | 4.803012e-05 | 28.000 | 5.718910 | 5.718885 | 4.363799e-06 |
| 6.000 | 2.576501 | 2.576576 | 2.915267e-05 | 29.000 | 5.819385 | 5.819362 | 4.043973e-06 |
| 7.000 | 2.802106 | 2.802153 | 1.679971e-05 | 30.000 | 5.918067 | 5.918045 | 3.732856e-06 |
| 8.000 | 3.011157 | 3.011183 | 8.570528e-06 | 31.000 | 6.015052 | 6.015032 | 3.431293e-06 |
| 9.000 | 3.206358 | 3.206367 | 3.017294e-06 | 32.000 | 6.110427 | 6.110408 | 3.139839e-06 |
| 10.000 | 3.389851 | 3.389848 | 7.543035e-07 | 33.000 | 6.204273 | 6.204255 | 2.858819e-06 |
| 11.000 | 3.563340 | 3.563328 | 3.312759e-06 | 34.000 | 6.296662 | 6.296646 | 2.588387e-06 |
| 12.000 | 3.728189 | 3.728170 | 5.029516e-06 | 35.000 | 6.387663 | 6.387649 | 2.328562e-06 |
| 13.000 | 3.885500 | 3.885476 | 6.153370e-06 | 36.000 | 6.477340 | 6.477326 | 2.079262e-06 |
| 14.000 | 4.036175 | 4.036147 | 6.854634e-06 | 37.000 | 6.565749 | 6.565737 | 1.840329e-06 |
| 15.000 | 4.180960 | 4.180930 | 7.252098e-06 | 38.000 | 6.652945 | 6.652934 | 1.611547e-06 |
| 16.000 | 4.320479 | 4.320447 | 7.429988e-06 | 39.000 | 6.738979 | 6.738969 | 1.392660e-06 |
| 17.000 | 4.455258 | 4.455225 | 7.448851e-06 | 40.000 | 6.823897 | 6.823888 | 1.183380e-06 |
| 18.000 | 4.585743 | 4.585710 | 7.352732e-06 | 42.000 | 6.990558 | 6.990552 | 7.924104e-07 |
| 19.000 | 4.712320 | 4.712286 | 7.173995e-06 | 46.000 | 7.312242 | 7.312241 | 1.117540e-07 |
| 20.000 | 4.835319 | 4.835285 | 6.936606e-06 | 48.000 | 7.467794 | 7.467795 | 1.830904e-07 |
| 21.000 | 4.955029 | 4.954996 | 6.658439e-06 | 50.000 | 7.620123 | 7.620127 | 4.508943e-07 |
| 22.000 | 5.071705 | 5.071673 | | | | | |

Table A17: Comparison between RK and MC solutions of $\theta(\xi)$ with polytrope index $n = -4$.

| $\xi$ | $\theta_{RK}$ | $\theta_{MC}$ | Relative error | $\xi$ | $\theta_{RK}$ | $\theta_{MC}$ | Relative error |
|---|---|---|---|---|---|---|---|
| 0.000 | 1.00000e+00 | 1.00000e+00 | 0.000000e+00 | 23.000 | 3.964045 | 3.964010 | 8.814443e-06 |
| 1.000 | 1.140660 | 1.140885 | 0.0001976200 | 24.000 | 4.031416 | 4.031382 | 8.235174e-06 |
| 2.000 | 1.406659 | 1.406874 | 0.0001526830 | 25.000 | 4.097031 | 4.097000 | 7.673072e-06 |
| 3.000 | 1.665324 | 1.665474 | 9.014820e-05 | 26.000 | 4.161012 | 4.160983 | 7.130028e-06 |
| 4.000 | 1.893088 | 1.893182 | 4.954760e-05 | 27.000 | 4.223467 | 4.223439 | 6.607193e-06 |
| 5.000 | 2.092495 | 2.092547 | 2.490940e-05 | 28.000 | 4.284491 | 4.284465 | 6.105170e-06 |
| 6.000 | 2.269121 | 2.269143 | 9.820040e-06 | 29.000 | 4.344173 | 4.344148 | 5.624158e-06 |
| 7.000 | 2.427727 | 2.427728 | 4.281099e-07 | 30.000 | 4.402590 | 4.402567 | 5.164062e-06 |
| 8.000 | 2.571934 | 2.571920 | 5.468375e-06 | 31.000 | 4.459814 | 4.459793 | 4.724572e-06 |
| 9.000 | 2.704447 | 2.704422 | 9.154824e-06 | 32.000 | 4.515911 | 4.515892 | 4.305225e-06 |
| 10.000 | 2.827304 | 2.827272 | 1.140705e-05 | 33.000 | 4.570940 | 4.570922 | 3.905450e-06 |
| 11.000 | 2.942067 | 2.942030 | 1.270903e-05 | 34.000 | 4.624955 | 4.624939 | 3.524605e-06 |
| 12.000 | 3.049948 | 3.049907 | 1.337174e-05 | 35.000 | 4.678007 | 4.677993 | 3.162004e-06 |
| 13.000 | 3.151907 | 3.151864 | 1.360011e-05 | 36.000 | 4.730142 | 4.730129 | 2.816932e-06 |
| 14.000 | 3.248715 | 3.248671 | 1.353210e-05 | 37.000 | 4.781403 | 4.781391 | 2.488667e-06 |
| 15.000 | 3.340998 | 3.340954 | 1.326227e-05 | 38.000 | 4.831829 | 4.831818 | 2.176485e-06 |
| 16.000 | 3.429273 | 3.429229 | 1.285638e-05 | 39.000 | 4.881456 | 4.881447 | 1.879671e-06 |
| 17.000 | 3.513971 | 3.513927 | 1.236070e-05 | 40.000 | 4.930319 | 4.930311 | 1.597526e-06 |
| 18.000 | 3.595453 | 3.595411 | 1.180805e-05 | 42.000 | 5.025878 | 5.025873 | 1.074542e-06 |
| 19.000 | 3.674030 | 3.673989 | 1.122191e-05 | 44.000 | 5.118735 | 5.118732 | 6.023595e-07 |
| 20.000 | 3.749963 | 3.749924 | 1.061906e-05 | 46.000 | 5.209092 | 5.209091 | 1.762041e-07 |
| 21.000 | 3.823483 | 3.823444 | 1.001155e-05 | 48.000 | 5.297126 | 5.297127 | 2.082914e-07 |
| 22.000 | 3.894785 | 3.894749 | 9.407985e-06 | 50.000 | 5.382995 | 5.382998 | 5.550936e-07 |

Table A18: Comparison between RK and MC solutions of $\theta(\xi)$ with polytrope index $n = -5$.

| $\xi$ | $\theta_{RK}$ | $\theta_{MC}$ | Relative error | $\xi$ | $\theta_{RK}$ | $\theta_{MC}$ | Relative error |
|---|---|---|---|---|---|---|---|
| 0.000 | 1.00000e+00 | 1.00000e+00 | 0.000000e+00 | 23.000 | 3.278878 | 3.278846 | 9.810960e-06 |
| 1.000 | 1.135601 | 1.135808 | 0.0001823063 | 24.000 | 3.324933 | 3.324903 | 9.088937e-06 |
| 2.000 | 1.373127 | 1.373300 | 0.0001265409 | 25.000 | 3.369666 | 3.369638 | 8.397541e-06 |
| 3.000 | 1.589799 | 1.589906 | 6.685255e-05 | 26.000 | 3.413172 | 3.413146 | 7.737074e-06 |
| 4.000 | 1.772978 | 1.773033 | 3.081464e-05 | 27.000 | 3.455535 | 3.455510 | 7.107333e-06 |
| 5.000 | 1.928973 | 1.928992 | 9.939410e-06 | 28.000 | 3.496829 | 3.496806 | 6.507764e-06 |
| 6.000 | 2.064361 | 2.064356 | 2.301652e-06 | 29.000 | 3.537121 | 3.537100 | 5.937573e-06 |
| 7.000 | 2.184023 | 2.184003 | 9.543121e-06 | 30.000 | 3.576474 | 3.576455 | 5.395802e-06 |
| 8.000 | 2.291429 | 2.291398 | 1.378873e-05 | 31.000 | 3.614942 | 3.614924 | 4.881394e-06 |
| 9.000 | 2.389067 | 2.389029 | 1.618169e-05 | 32.000 | 3.652575 | 3.652559 | 4.393229e-06 |
| 10.000 | 2.478760 | 2.478717 | 1.740021e-05 | 33.000 | 3.689419 | 3.689405 | 3.930164e-06 |
| 11.000 | 2.561872 | 2.561826 | 1.786125e-05 | 34.000 | 3.725516 | 3.725503 | 3.491051e-06 |
| 12.000 | 2.639447 | 2.639400 | 1.782897e-05 | 35.000 | 3.760904 | 3.760892 | 3.074753e-06 |
| 13.000 | 2.712298 | 2.712251 | 1.747503e-05 | 36.000 | 3.795617 | 3.795607 | 2.680158e-06 |
| 14.000 | 2.781072 | 2.781025 | 1.691324e-05 | 37.000 | 3.829690 | 3.829681 | 2.306190e-06 |
| 15.000 | 2.846290 | 2.846244 | 1.622031e-05 | 38.000 | 3.863151 | 3.863144 | 1.951805e-06 |
| 16.000 | 2.908376 | 2.908331 | 1.544854e-05 | 39.000 | 3.896029 | 3.896023 | 1.616007e-06 |
| 17.000 | 2.96783 | 2.967639 | 1.463393e-05 | 40.000 | 3.925142 | 3.925137 | 1.328887e-06 |
| 18.000 | 3.024504 | 3.024462 | 1.380132e-05 | 42.000 | 3.991411 | 3.991408 | 7.107817e-0 |
| 19.000 | 3.079088 | 3.079048 | 1.296788e-05 | 44.000 | 4.052510 | 4.052509 | 1.838675e-07 |
| 20.000 | 3.131648 | 3.131610 | 1.214540e-05 | 46.000 | 4.111796 | 4.111798 | 2.890026e-07 |
| 21.000 | 3.182364 | 3.182328 | 1.134188e-05 | 48.000 | 4.169405 | 4.169408 | 7.132958e-07 |
| 22.000 | 3.231395 | 3.231361 | 1.056260e-05 | 50.000 | 4.225452 | 4.225457 | 1.093902e-06 |

Table A19: Comparison between RK and MC solutions of $\theta(\xi)$ with polytrope index $n = -10$.

| $\xi$ | $\theta_{RK}$ | $\theta_{MC}$ | Relative error | $\xi$ | $\theta_{RK}$ | $\theta_{MC}$ | Relative error |
|---|---|---|---|---|---|---|---|
| 0.000 | 1.00000e+00 | 1.00000e+00 | 0.000000e+00 | 23.000 | 2.044024 | 2.044007 | 7.960936e-06 |
| 1.000 | 1.115647 | 1.115789 | 0.0001275456 | 24.000 | 2.059288 | 2.059274 | 7.068898e-06 |
| 2.000 | 1.268750 | 1.268821 | 5.574577e-05 | 25.000 | 2.074032 | 2.074019 | 6.231806e-06 |
| 3.000 | 1.383919 | 1.383936 | 1.224969e-05 | 26.000 | 2.088295 | 2.088283 | 5.446560e-06 |
| 4.000 | 1.471326 | 1.471313 | 8.860233e-06 | 27.000 | 2.102112 | 2.102102 | 4.710112e-06 |
| 5.000 | 1.540827 | 1.540798 | 1.908105e-05 | 28.000 | 2.115514 | 2.115505 | 4.019506e-06 |
| 6.000 | 1.598300 | 1.598262 | 2.385867e-05 | 29.000 | 2.128530 | 2.128523 | 3.371933e-06 |
| 7.000 | 1.647272 | 1.647230 | 2.576803e-05 | 30.000 | 2.141184 | 2.141178 | 2.764725e-06 |
| 8.000 | 1.689971 | 1.689927 | 2.610454e-05 | 31.000 | 2.153500 | 2.153495 | 2.195354e-06 |
| 9.000 | 1.727871 | 1.727827 | 2.556004e-05 | 32.000 | 2.165498 | 2.165494 | 1.661444e-06 |
| 10.000 | 1.761992 | 1.761949 | 2.452301e-05 | 33.000 | 2.177196 | 2.177194 | 1.160765e-06 |
| 11.000 | 1.793067 | 1.793025 | 2.321937e-05 | 34.000 | 2.188612 | 2.188611 | 6.912254e-07 |
| 12.000 | 1.821633 | 1.821594 | 2.178431e-05 | 35.000 | 2.199761 | 2.199761 | 2.508766e-07 |
| 13.000 | 1.848101 | 1.848064 | 2.030003e-05 | 36.000 | 2.210657 | 2.210657 | 1.621052e-07 |
| 14.000 | 1.872788 | 1.872753 | 1.881678e-05 | 37.000 | 2.221313 | 2.221314 | 5.494215e-07 |
| 15.000 | 1.895942 | 1.895909 | 1.736519e-05 | 38.000 | 2.231741 | 2.231743 | 9.126640e-07 |
| 16.000 | 1.917766 | 1.917735 | 1.596349e-05 | 39.000 | 2.241952 | 2.241955 | 1.253311e-06 |
| 17.000 | 1.938421 | 1.938392 | 1.462206e-05 | 40.000 | 2.251957 | 2.251960 | 1.572748e-06 |
| 18.000 | 1.958042 | 1.958016 | 1.334617e-05 | 42.000 | 2.271383 | 2.271388 | 2.153063e-06 |
| 19.000 | 1.994616 | 1.994594 | 1.099696e-05 | 44.000 | 2.290089 | 2.290095 | 2.662931e-06 |
| 20.000 | 1.994616 | 1.994594 | 1.099696e-05 | 46.000 | 2.308132 | 2.308139 | 3.110478e-06 |
| 21.000 | 2.011743 | 2.011723 | 9.922110e-06 | 48.000 | 2.325565 | 2.325573 | 3.502799e-06 |
| 22.000 | 2.028192 | 2.028174 | 8.911012e-06 | 50.000 | 2.342434 | 2.342443 | 3.846096e-06 |

Table A20: Comparison between RK and MC solutions of $\theta(\xi)$ with polytrope index $n \to \infty$.

| $\xi$ | $\theta_{RK}$ | $\theta_{MC}$ | Relative error | $\xi$ | $\theta_{RK}$ | $\theta_{MC}$ | Relative error |
|---|---|---|---|---|---|---|---|
| 0.000 | 0.00000e+00 | 0.00000e+00 | 0.000000e+00 | 18.000 | 5.162940 | 5.162785 | 2.997646e-05 |
| 0.500 | 0.04116308 | 0.04132496 | 0.003932624 | 19.000 | 5.287224 | 5.287063 | 3.040928e-05 |
| 1.000 | 0.1588357 | 0.1591336 | 0.001875320 | 20.000 | 5.403886 | 5.403721 | 3.062334e-05 |
| 2.000 | 0.5598280 | 0.5602713 | 0.0007917708 | 21.000 | 5.513725 | 5.513556 | 3.066668e-05 |
| 3.000 | 1.063337 | 1.063782 | 0.0004186039 | 22.000 | 5.617429 | 5.617258 | 3.057622e-05 |
| 4.000 | 1.572233 | 1.572610 | 0.0002401128 | 23.000 | 5.715594 | 5.715420 | 3.038069e-05 |
| 5.000 | 2.044091 | 2.044382 | 0.0001425249 | 24.000 | 5.808736 | 5.808561 | 3.010262e-05 |
| 6.000 | 2.467208 | 2.467417 | 8.462947e-05 | 25.000 | 5.897309 | 5.897134 | 2.975982e-05 |
| 7.000 | 2.842587 | 2.842724 | 4.820993e-05 | 26.000 | 5.981712 | 5.981536 | 2.936645e-05 |
| 8.000 | 3.175394 | 3.175472 | 2.426887e-05 | 27.000 | 6.062294 | 6.062118 | 2.893384e-05 |
| 9.000 | 3.471556 | 3.471584 | 7.994843e-06 | 28.000 | 6.139367 | 6.139192 | 2.847110e-05 |
| 10.000 | 3.736558 | 3.736545 | 3.350720e-06 | 29.000 | 6.213207 | 6.213033 | 2.798560e-05 |
| 11.000 | 3.975117 | 3.975072 | 1.140709e-05 | 30.000 | 6.284060 | 6.283888 | 2.748330e-05 |
| 12.000 | 4.191176 | 4.191104 | 1.719919e-05 | 31.000 | 6.352149 | 6.351978 | 2.696906e-05 |
| 13.000 | 4.387992 | 4.387898 | 2.139243e-05 | 32.000 | 6.417671 | 6.417502 | 2.644681e-05 |
| 14.000 | 4.568251 | 4.568140 | 2.443182e-05 | 33.000 | 6.480806 | 6.480638 | 2.591979e-05 |
| 15.000 | 4.734176 | 4.734050 | 2.662318e-05 | 34.000 | 6.541714 | 6.541548 | 2.539062e-05 |
| 16.000 | 4.887613 | 4.887475 | 2.818173e-05 | 35.000 | 6.600543 | 6.600379 | 2.486147e-05 |
| 17.000 | 5.030102 | 5.029955 | 2.926188e-05 | | | | |

Table A21: Comparison between RK and MC solutions of $\psi(\eta)$ with constants $C = 0.01$ & $0.1$.

| | $C = 0.01$ | | | | $C = 0.1$ | | |
|---|---|---|---|---|---|---|---|
| $\eta$ | $\theta_{RK}$ | $\theta_{MC}$ | Relative error | $\eta$ | $\theta_{RK}$ | $\theta_{MC}$ | Relative error |
| 0.000 | 1.00000e+00 | 1.00000e+00 | 0.000000e+00 | 0.000 | 1.00000e+00 | 1.00000e+00 | 0.000000e+00 |
| 0.500 | 0.9604334 | 0.9602827 | 0.0001568452 | 0.500 | 0.9656478 | 0.9655167 | 1.357327e-04 |
| 1.000 | 0.8571405 | 0.8569046 | 0.0002752544 | 1.000 | 0.8754712 | 0.8752640 | 0.0002366674 |
| 1.500 | 0.7234100 | 0.7231662 | 0.0003370373 | 1.500 | 0.7578572 | 0.7576410 | 0.0002852371 |
| 2.000 | 0.5885526 | 0.5883501 | 0.0003441379 | 2.000 | 0.6387512 | 0.6385709 | 0.0002822953 |
| 2.500 | 0.4684922 | 0.4683486 | 0.0003065139 | 2.500 | 0.5330399 | 0.5329130 | 0.0002380624 |
| 3.000 | 0.3681576 | 0.3680721 | 0.0002323289 | 3.000 | 0.4458162 | 0.4457430 | 0.0001641943 |
| 3.500 | 0.2867141 | 0.2866786 | 1.238057e-04 | 3.300 | 0.4022845 | 0.4022404 | 1.098144e-04 |
| 4.000 | 0.2212402 | 0.2212455 | 2.435932e-05 | 3.550 | 0.3705892 | 0.3705668 | 6.027593e-05 |
| 4.200 | 0.1987934 | 0.1988129 | 9.769646e-05 | 3.600 | 0.3647122 | 0.3646939 | 4.998560e-05 |
| 4.400 | 0.1781920 | 0.1782243 | 0.0001812103 | 3.650 | 0.3589814 | 0.3589672 | 3.958061e-05 |
| 4.600 | 0.1592579 | 0.1593020 | 0.0002769787 | 3.700 | 0.3533934 | 0.3533831 | 2.906585e-05 |
| 4.800 | 0.1418276 | 0.1418826 | 0.0003877678 | 3.750 | 0.3479447 | 0.3479382 | 1.844580e-05 |
| 5.000 | 0.1257519 | 0.1258169 | 0.0005173050 | 3.800 | 0.3426318 | 0.3426291 | 7.724584e-06 |
| 5.100 | 0.1181791 | 0.1182489 | 0.0005906332 | 3.900 | 0.3323994 | 0.3324041 | 1.400726e-05 |
| 5.200 | 0.1108950 | 0.1109694 | 0.0006707111 | 4.000 | 0.3226680 | 0.3226796 | 3.610706e-05 |
| 5.357 | 0.1000059 | 0.10008711 | 0.0008124600 | 4.069 | 0.3162301 | 0.3162464 | 5.156471e-05 |

Table A22: Comparison between RK and MC solutions of $\psi(\eta)$ with constants $C = 0.2$ & $0.3$.

| | $C = 0.2$ | | | | $C = 0.3$ | | |
|---|---|---|---|---|---|---|---|
| $\eta$ | $\theta_{RK}$ | $\theta_{MC}$ | Relative error | $\eta$ | $\theta_{RK}$ | $\theta_{MC}$ | Relative error |
| 0.000 | 1.00000e+00 | 1.00000e+00 | 0.000000e+00 | 0.000 | 1.00000e+00 | 1.00000e+00 | 0.000000e+00 |
| 0.500 | 0.9711551 | 0.9710448 | 1.136257e-04 | 0.500 | 0.9763417 | 0.9762509 | 9.295675e-05 |
| 1.000 | 0.8949289 | 0.8947525 | 0.0001971433 | 1.000 | 0.91336268 | 0.9132157 | 0.000160887 |
| 1.500 | 0.7945548 | 0.7943683 | 0.0002346878 | 1.500 | 0.8295005 | 0.8293428 | 0.0001902091 |
| 2.000 | 0.6921701 | 0.6920132 | 0.0002267944 | 2.000 | 0.7430857 | 0.7429509 | 0.0001813331 |
| 2.100 | 0.6728038 | 0.6726554 | 0.0002205926 | 2.100 | 0.7266719 | 0.7265442 | 0.0001757375 |
| 2.200 | 0.6539855 | 0.6538461 | 0.0002130886 | 2.200 | 0.7107107 | 0.7105906 | 0.0001690950 |
| 2.300 | 0.6357639 | 0.6356340 | 0.0002043915 | 2.300 | 0.6952487 | 0.6951364 | 0.0001615106 |
| 2.400 | 0.6181762 | 0.6180559 | 0.0001946114 | 2.400 | 0.6803224 | 0.6802182 | 0.0001530905 |
| 2.500 | 0.6012488 | 0.6011382 | 0.0001838574 | 2.500 | 0.6659590 | 0.6658631 | 0.0001439404 |
| 2.600 | 0.5849989 | 0.5848982 | 0.0001722373 | 2.600 | 0.6521774 | 0.6520899 | 0.0001341642 |
| 2.700 | 0.5694358 | 0.5693448 | 0.0001598560 | 2.700 | 0.6389889 | 0.6389097 | 0.0001238630 |
| 2.800 | 0.5545614 | 0.5544800 | 0.0001468156 | 2.800 | 0.6263983 | 0.6263274 | 1.131343e-04 |
| 2.900 | 0.5403719 | 0.5402999 | 0.0001332138 | 2.900 | 0.6144044 | 0.6143417 | 1.020706e-04 |
| 3.000 | 0.5268581 | 0.5267953 | 0.0001191438 | 3.000 | 0.6030012 | 0.6029465 | 9.075923e-05 |
| 3.200 | 0.5018003 | 0.5017551 | 8.994681e-05 | 3.100 | 0.5921780 | 0.5921310 | 7.928149e-05 |
| 3.400 | 0.4792399 | 0.4792113 | 5.985975e-05 | 3.200 | 0.5819203 | 0.5818809 | 6.771207e-05 |
| 3.500 | 0.4688382 | 0.4688173 | 4.465261e-05 | 3.300 | 0.5722101 | 0.5721780 | 5.611871e-05 |
| 3.600 | 0.4589869 | 0.4589734 | 2.941249e-05 | 3.400 | 0.5630267 | 0.5630016 | 4.456192e-05 |
| 3.700 | 0.4496568 | 0.4496504 | 1.418607e-05 | 3.500 | 0.5543459 | 0.5543276 | 3.309489e-05 |
| 3.727 | 0.4472230 | 0.4472185 | 1.008237e-05 | 3.580 | 0.5477452 | 0.5477320 | 2.401760e-05 |

Table A23: Comparison between RK and MC solutions of $\psi(\eta)$ with constants $C = 0.4$ & $0.5$.

| | $C = 0.4$ | | | | $C = 0.5$ | | |
|---|---|---|---|---|---|---|---|
| $\eta$ | $\theta_{RK}$ | $\theta_{MC}$ | Relative error | $\eta$ | $\theta_{RK}$ | $\theta_{MC}$ | Relative error |
| 0.000 | 1.00000e+00 | 1.00000e+00 | 0.000000e+00 | 0.000 | 1.00000e+00 | 1.00000e+00 | 0.000000e+00 |
| 0.500 | 0.9811835 | 0.9811111 | 7.379295e-05 | 0.500 | 0.9856505 | 0.9855952 | 5.620388e-05 |
| 1.000 | 0.9306870 | 0.9305681 | 0.0001277466 | 1.000 | 0.9467932 | 0.9467008 | 9.761168e-05 |
| 1.500 | 0.8625650 | 0.8624350 | 0.0001508094 | 1.500 | 0.8935688 | 0.8934654 | 0.0001157336 |
| 2.000 | 0.7914187 | 0.7913053 | 0.0001431995 | 2.000 | 0.8370121 | 0.8369196 | 0.0001104989 |
| 2.100 | 0.7778116 | 0.7777038 | 0.0001386363 | 2.100 | 0.8260864 | 0.8259979 | 0.0001071285 |
| 2.200 | 0.7645556 | 0.7644537 | 0.0001332529 | 2.200 | 0.81541031 | 0.8153262 | .031329e-04 |
| 2.300 | 0.7516939 | 0.7515983 | 0.0001271400 | 2.300 | 0.8050221 | 0.8049428 | 9.858252e-05 |
| 2.400 | 0.7392613 | 0.7391723 | 0.0001203888 | 2.400 | 0.7949535 | 0.7948792 | 9.354869e-05 |
| 2.500 | 0.7272846 | 0.7272023 | 0.0001130903 | 2.500 | 0.7852301 | 0.7851609 | 8.810257e-05 |
| 2.600 | 0.7157837 | 0.7157083 | 1.053338e-04 | 2.600 | 0.7758717 | 0.7758078 | 8.231407e-05 |
| 2.700 | 0.7047717 | 0.7047032 | 9.720561e-05 | 2.700 | 0.7668926 | 0.7668342 | 7.625095e-05 |
| 2.800 | 0.6942560 | 0.6941944 | 8.878862e-05 | 2.800 | 0.7583024 | 0.7582493 | 6.997804e-05 |
| 2.900 | 0.6842384 | 0.6841836 | 8.016108e-05 | 2.900 | 0.7501059 | 0.7500582 | 6.355666e-05 |
| 3.000 | 0.6747162 | 0.6746680 | 7.139613e-05 | 3.000 | 0.7423039 | 0.7422615 | 5.704401e-05 |
| 3.100 | 0.6656822 | 0.6656406 | 6.256125e-05 | 3.100 | 0.7348934 | 0.7348563 | 5.049281e-05 |
| 3.200 | 0.6571257 | 0.6570904 | 5.371790e-05 | 3.200 | 0.7278680 | 0.7278360 | 4.395104e-05 |
| 3.300 | 0.6490323 | 0.6490032 | 4.492129e-05 | 3.300 | 0.7212182 | 0.7211912 | 3.746185e-05 |
| 3.400 | 0.6413845 | 0.6413613 | 3.622043e-05 | 3.400 | 0.7149314 | 0.7149092 | 3.106369e-05 |
| 3.500 | 0.6341614 | 0.6341439 | 2.765903e-05 | 3.500 | 0.7089920 | 0.7089744 | 2.479118e-05 |
| 3.524 | 0.6324884 | 0.6324722 | 2.562977e-05 | 3.532 | 0.7071617 | 0.7071456 | 2.281613e-05 |

Table A24: Comparison between RK and MC solutions of $\psi(\eta)$ with constants $C = 0.6$ & $0.7$.

| | $C = 0.6$ | | | | $C = 0.7$ | | |
|---|---|---|---|---|---|---|---|
| $\eta$ | $\theta_{RK}$ | $\theta_{MC}$ | Relative error | $\eta$ | $\theta_{RK}$ | $\theta_{MC}$ | Relative error |
| 0.000 | 1.00000e+00 | 1.00000e+00 | 0.000000e+00 | 0.000 | 1.00000e+00 | 1.00000e+00 | 0.000000e+00 |
| 0.500 | 0.9897043 | 0.9896644 | 4.031136e-05 | 0.500 | 0.9932918 | 0.9932657 | 2.628113e-05 |
| 1.000 | 0.9615379 | 0.9614701 | 7.047888e-05 | 1.000 | 0.9747213 | 0.9746760 | 4.645837e-05 |
| 1.500 | 0.9222581 | 0.9221802 | 8.446162e-05 | 1.500 | 0.9482583 | 0.9482045 | 5.670576e-05 |
| 2.000 | 0.8795931 | 0.8795210 | 8.189307e-05 | 2.000 | 0.9186996 | 0.9186477 | 5.646971e-05 |
| 2.100 | 0.8712368 | 0.8711674 | 7.970403e-05 | 2.100 | 0.9128021 | 0.9127516 | 5.532516e-05 |
| 2.200 | 0.8630349 | 0.8629684 | 7.705819e-05 | 2.200 | 0.9069769 | 0.9069281 | 5.387246e-05 |
| 2.300 | 0.8550190 | 0.8549557 | 7.400443e-05 | 2.300 | 0.9012477 | 0.9012007 | 5.214033e-05 |
| 2.400 | 0.8472164 | 0.8471566 | 7.059263e-05 | 2.400 | 0.8956356 | 0.8955907 | 5.015853e-05 |
| 2.500 | 0.8396500 | 0.8395938 | 6.687291e-05 | 2.500 | 0.8901590 | 0.8901163 | 4.795735e-05 |
| 2.600 | 0.8472164 | 0.8471566 | 7.059263e-05 | 2.600 | 0.8848337 | 0.8847933 | 4.556717e-05 |
| 2.700 | 0.8252958 | 0.8252473 | 5.870661e-05 | 2.700 | 0.8796728 | 0.8796349 | 4.301802e-05 |
| 2.800 | 0.8185334 | 0.8184889 | 5.435492e-05 | 2.800 | 0.8746870 | 0.8746517 | 4.033921e-05 |
| 2.900 | 0.8120585 | 0.8120179 | 4.988395e-05 | 2.900 | 0.8698847 | 0.8698520 | 3.755899e-05 |
| 3.000 | 0.8058748 | 0.8058383 | 4.533529e-05 | 3.000 | 0.8652719 | 0.8652419 | 3.470429e-05 |
| 3.100 | 0.7999836 | 0.7999510 | 4.074750e-05 | 3.100 | 0.8608526 | 0.8608252 | 3.180048e-05 |
| 3.200 | 0.7943828 | 0.7943541 | 3.615595e-05 | 3.200 | 0.8566287 | 0.8566040 | 2.887121e-05 |
| 3.300 | 0.7890683 | 0.7890433 | 3.159269e-05 | 3.300 | 0.8526004 | 0.8525783 | 2.593823e-05 |
| 3.400 | 0.7840330 | 0.7840117 | 2.708650e-05 | 3.400 | 0.8487660 | 0.8487465 | 2.302140e-05 |
| 3.500 | 0.7792678 | 0.7792502 | 2.266316e-05 | 3.500 | 0.8451223 | 0.8451053 | 2.013865e-05 |
| 3.603 | 0.7746296 | 0.7746155 | 1.821920e-05 | 3.600 | 0.8416643 | 0.8416497 | 1.730615e-05 |
| | | | | 3.700 | 0.8383855 | 0.8383733 | 1.453868e-05 |
| | | | | 3.750 | 0.8366865 | 0.8366756 | 1.307657e-05 |

Table A25: Comparison between RK and MC solutions of $\psi(\eta)$ with constant $C = 0.8$.

| $\eta$ | $\theta_{RK}$ | $\theta_{MC}$ | Relative error | $\eta$ | $\theta_{RK}$ | $\theta_{MC}$ | Relative error |
|---|---|---|---|---|---|---|---|
| 0.000 | 1.00000e+00 | 1.00000e+00 | 0.000000e+00 | 2.100 | 0.9499440 | 0.9499123 | 3.337009e-05 |
| 0.100 | 0.9998511 | 0.9998481 | 2.984004e-06 | 2.200 | 0.9463841 | 0.9463530 | 3.284562e-05 |
| 0.200 | 0.9994054 | 0.9993994 | 5.932689e-06 | 2.300 | 0.9428509 | 0.9428205 | 3.215834e-05 |
| 0.300 | 0.9986665 | 0.9986576 | 8.832321e-06 | 2.400 | 0.9393579 | 0.9393284 | 3.132147e-05 |
| 0.400 | 0.9976403 | 0.9976287 | 1.165809e-05 | 2.500 | 0.9359173 | 0.9358889 | 3.034882e-05 |
| 0.500 | 0.9963349 | 0.9963206 | 1.438666e-05 | 2.600 | 0.9325403 | 0.9325130 | 2.925458e-05 |
| 0.600 | 0.9947606 | 0.9947437 | 1.699611e-05 | 2.700 | 0.9292364 | 0.9292103 | 2.805312e-05 |
| 0.700 | 0.9929295 | 0.9929102 | 1.946626e-05 | 2.800 | 0.9260143 | 0.9259895 | 2.675880e-05 |
| 0.800 | 0.9908555 | 0.9908340 | 2.177879e-05 | 2.900 | 0.9228813 | 0.9228579 | 2.538583e-05 |
| 0.900 | 0.9885541 | 0.9885305 | 2.391751e-05 | 3.000 | 0.9198434 | 0.9198214 | 2.394805e-05 |
| 1.000 | 0.9860421 | 0.9860166 | 2.586848e-05 | 3.100 | 0.9169057 | 0.9168851 | 2.245886e-05 |
| 1.100 | 0.9833374 | 0.9833102 | 2.762015e-05 | 3.200 | 0.9140719 | 0.9140528 | 2.093106e-05 |
| 1.200 | 0.9804587 | 0.9804301 | 2.916340e-05 | 3.300 | 0.9113449 | 0.9113272 | 1.937679e-05 |
| 1.300 | 0.9774255 | 0.9773957 | 3.049166e-05 | 3.400 | 0.9087264 | 0.9087102 | 1.780740e-05 |
| 1.400 | 0.9742574 | 0.9742266 | 3.160084e-05 | 3.500 | 0.9062172 | 0.9062025 | 1.623347e-05 |
| 1.500 | 0.9709744 | 0.9709428 | 3.248928e-05 | 3.600 | 0.9038173 | 0.9038041 | 1.466473e-05 |
| 1.600 | 0.9675963 | 0.9675642 | 3.315768e-05 | 3.700 | 0.9015257 | 0.9015139 | 1.311005e-05 |
| 1.700 | 0.9641427 | 0.9641103 | 3.360897e-05 | 3.800 | 0.8993406 | 0.8993302 | 1.157751e-05 |
| 1.800 | 0.9606326 | 0.9606001 | 3.384814e-05 | 3.900 | 0.8972594 | 0.8972504 | 1.007446e-05 |
| 1.900 | 0.9570845 | 0.9570521 | 3.388207e-05 | 4.045 | 0.8944189 | 0.8944118 | 7.961630e-06 |
| 2.000 | 0.9535161 | 0.9534839 | 3.371936e-05 | | | | |